\documentclass[fleqn,usenatbib]{mnras}

\usepackage{newtxtext,newtxmath}

\usepackage[T1]{fontenc}
\usepackage{ae,aecompl}

\usepackage{xcolor}

\usepackage[utf8]{inputenc}
\usepackage[T1]{fontenc}

\usepackage{graphicx}	
\usepackage{amsmath}	
\usepackage{ulem}

\newcommand{\RmatEl}{\mathsf{R}}
\newcommand{\Rmat}{\boldsymbol{\RmatEl}}
\newcommand{\RmatLazyEl}{\mathsf{R}^\daleth}
\newcommand{\RmatLazy}{\boldsymbol{\RmatLazyEl}}

\newcommand{\xvec}{{\bf x}}

\newcommand{\phatvec}{\boldsymbol{\hat{p}}}

\newcommand{\phat}{\hat{p}}
\newcommand{\MmatEl}{\mathsf{M}}
\newcommand{\Mmat}{\boldsymbol{\MmatEl}}
\newcommand{\MmatIdEl}{\MmatEl_\text{ID}}

\newcommand{\MmatId}{\boldsymbol{\MmatIdEl}}

\newcommand{\SigmaMatHat}{\boldsymbol{\hat{\Sigma}}}
\newcommand{\SigmaMatHatEl}{\hat{\Sigma}}

\newcommand{\Fmat}{\boldsymbol{\mathsf{F}}}

\newcommand{\Welt}{\mathsf{W}}
\newcommand{\Wmat}{\boldsymbol{\mathsf{{W}}}}
\newcommand{\WmatHat}{\boldsymbol{\hat{\mathsf{W}}}}
\newcommand{\Lmat}{\boldsymbol{\mathsf{L}}}
\newcommand{\Imat}{\boldsymbol{\mathsf{I}}}
\newcommand{\xcovMatHat}{\boldsymbol{\hat{\mathsf{C}}}}
\newcommand{\dxcovMatAlpha}{\boldsymbol{{\mathsf{C}_{,\alpha}}}}
\newcommand{\dxcovMatAlphaHat}{\boldsymbol{{\widehat{\mathsf{C}}_{,\alpha}}}}
\newcommand{\dxcovMatBetaHat}{\boldsymbol{{\widehat{\mathsf{C}}_{,\beta}}}}

\newcommand{\dxcovMatBeta}{\boldsymbol{\mathsf{C}_{,\beta}}}
\newcommand{\dxcovMatGamma}{\boldsymbol{\mathsf{C}_{,\gamma}}}

\newcommand{\Hmat}{\boldsymbol{\mathsf{H}}}

\newcommand{\xcovMatEl}{\mathsf{C}}
\newcommand{\xcovMat}{\boldsymbol{\xcovMatEl}}
\newcommand{\ftxcovMatEl}{\widetilde{\mathsf{C}}}

\newcommand{\RmatRestEl}{\RmatEl^\text{REST}}
\newcommand{\RmatRest}{\boldsymbol{\RmatRestEl}}
\newcommand{\HmatHat}{\boldsymbol{\hat{\mathsf{H}}}}
\newcommand{\HmatHatEl}{\hat{\mathsf{H}}}

\newcommand{\ftxcovMatLazyEl}{\widetilde{\mathsf{C}}^\daleth}
\newcommand{\ftxcovMatLazy}{\boldsymbol{\ftxcovMatLazyEl}}

\newcommand{\xcovMatDFTEl}{\xcovMatEl^\text{DFT}}
\newcommand{\xcovMatDFT}{\boldsymbol{\xcovMatDFTEl}}

\newcommand{\ftxcovMatDFTEl}{\ftxcovMatEl^\text{DFT}}

\newcommand{\xcovMatLazyEl}{\xcovMatEl^\daleth}
\newcommand{\xcovMatLazy}{\boldsymbol{\xcovMatLazyEl}}
\newcommand{\xcovMatLazyFGEl}{\xcovMatLazyEl_\text{FG}}
\newcommand{\xcovMatLazyFG}{\boldsymbol{\xcovMatLazyFGEl}}

\newcommand{\Ematalpha}{\boldsymbol{\mathsf{E}_{\alpha}}}
\newcommand{\Ematbeta}{\boldsymbol{\mathsf{E}_{\beta}}}
\newcommand{\Qmatalpha}{\boldsymbol{\mathsf{Q}_\alpha}}

\newcommand{\Ndata}{N_\text{d}}
\newcommand{\Nbp}{N_\text{b}}

\newcommand{\wmat}{\boldsymbol{\mathsf{w}}}
\newcommand{\Nmat}{\boldsymbol{\mathsf{N}}}
\newcommand{\Smat}{\boldsymbol{\mathsf{S}}}
\newcommand{\Amat}{\boldsymbol{\mathsf{A}}}
\newcommand{\nvec}{\boldsymbol{n}}
\newcommand{\fvec}{\boldsymbol{f}}
\newcommand{\svec}{\boldsymbol{s}}
\newcommand{\biasalpha}{b_\alpha}
\newcommand{\biasalphahat}{\hat{b}_\alpha}

\newcommand{\uvec}{{\bf u}}
\newcommand{\zvec}{{\bf z}}

\title[A Simple Foreground Filter]{{\tt DAYENU:} A Simple Filter of Smooth Foregrounds for Intensity Mapping Power Spectra}

\author[A. Ewall-Wice et al.]{Aaron Ewall-Wice,$^{1,2}$\thanks{E-mail: aaronew@berkeley.edu}
Nicholas Kern,$^{1}$
Joshua S. Dillon,$^{1,\dagger}$
Adrian Liu,$^{3}$ 
Aaron Parsons,$^{1}$
\newauthor
Saurabh Singh,$^{3}$
Adam Lanman,$^{4}$
Paul La Plante,$^{1,2}$
Nicolas Fagnoni,$^{5}$
\newauthor
Eloy de Lera Acedo,$^{5}$
David R. DeBoer,$^{1}$
Chuneeta Nunhokee,$^{1}$
Philip Bull,$^{6}$
\newauthor
Tzu-Ching Chang,$^{7,8}$
T.~Joseph~W.~Lazio,$^{7}$
James Aguirre,$^{9}$
Sean Weinberg,$^{10}$
\\
$^{1}$Department of Astronomy, University of California, Berkeley, CA 94720, USA\\
$^{2}$Berkeley Center for Cosmological Physics, University of California, Berkeley, CA 94720, USA\\
$^{3}$Department of Physics and McGill Space Institute, McGill University, 3600 University Street, Montreal, QC H3A 2T8, Canada\\
$^{4}$Department of Physics, Brown University, Providence, RI\\
$^{5}$Cavendish Astrophysics, University of Cambridge, Cambridge, UK\\
$^{6}$School of Physics \& Astronomy, Queen Mary University of London, London, UK\\
$^{7}$Jet Propulsion Laboratory, California Institute of Technology 4800 Oak Grove Dr, M/S 169-237, Pasadena CA 91109, USA\\
$^{8}$California Institute of Technology, 1200 E California Blvd, Pasadena, CA 91125, USA\\
$^{9}$Department of Physics and Astronomy, University of Pennsylvania, Philadelphia, PA\\
$^{10}$ QC Ware, Palo Alto, CA 94301, USA \\ 
$^{\dagger}$ National Science Foundational Astronomy and Astrophysics Postdoctoral Fellow
}

\date{Accepted XXX. Received YYY; in original form ZZZ}

\pubyear{2019}
\begin{document}
\label{firstpage}
\pagerange{\pageref{firstpage}--\pageref{lastpage}}
\maketitle

\begin{abstract}
We introduce {\tt DAYENU}, a linear, spectral filter for HI intensity mapping that achieves the desirable foreground mitigation and error minimization properties of inverse co-variance weighting with minimal modeling of the underlying data. Beyond 21\,cm power-spectrum estimation, our filter is suitable for any analysis where high dynamic-range removal of spectrally smooth foregrounds in irregularly (or regularly) sampled data is required, something required by many other intensity mapping techniques. Our filtering matrix is diagonalized  by Discrete Prolate Spheroidal Sequences which are an optimal basis to model band-limited foregrounds in 21\,cm intensity mapping experiments in the sense that they maximally concentrate power within a finite region of Fourier space. We show that {\tt DAYENU} enables the access of large-scale line-of-sight modes that are inaccessible to tapered DFT estimators. Since these modes have the largest SNRs, {\tt DAYENU} significantly increases the sensitivity of 21\,cm analyses over tapered Fourier transforms.  
 Slight modifications allow us to use DAYENU as a linear replacement for iterative delay CLEANing (DAYENUREST). We refer readers to the Code section at the end of this paper for links to examples and code.
\end{abstract}


\section{Introduction}\label{sec:intro}

Buried under vastly brighter foregrounds, redshifted 21\,cm emission from \ion{H}{i} at redshifts $z \gtrsim 6$ remains an elusive treasure trove of information on how the first stars and galaxies heated and subsequently ionized the universe. Experiments seeking to observe spatial 21\,cm fluctuations are attempting a first detection with the power spectrum statistic, $P(k)$ defined through, 
\begin{equation}
(2 \pi )^3 \delta^D\left({\bf k} - {\bf k^\prime} \right) P({\bf k}) = \langle \widetilde{T_b}({\bf k}) \widetilde{T_b}^*({\bf k^\prime}) \rangle - \langle \widetilde{T_b}({\bf k}) \rangle \langle \widetilde{T_b}^*({\bf k^\prime}) \rangle
\end{equation}
where $\delta^D$ is the Dirac delta-function, $T({\bf k})$ is the co-moving spatial Fourier transform of the cosmological brightness temperature field, 
\begin{equation}\label{eq:CONTINUOUSFOURIERTRANFORM}
\widetilde{T_b}({\bf k}) = \int d^3{\bf r} e^{i{\bf k} \cdot {\bf r}} T_b({\bf r}),
\end{equation}
and $\langle \cdot \rangle$ denotes an ensemble average. Gaussian random fields are completely described by the power-spectrum. The power spectrum is also a convenient statistic for non-Gaussian fields since we can take advantage of the fact that cosmological quantities approximtely obey statistical homogeneity and isotropy; allowing us to build sensitivity by averaging in spherical Fourier bins. 

Another convenient feature 21\,cm and other intensity mapping experiments is that foregrounds; which are expected to be intrinsically spectrally smooth, only occupy small wave-numbers along the line of sight (small $k_\parallel$) while 21\,cm and other spectral lines that trace cosmological structures have substantial fine-scale spectral features \citep{DiMatteo:2004, Datta:2010, Parsons:2012}. Thus, the native Fourier space of the power-spectrum is well-suited for performing foreground separation.

While single-dish experiments such as GBT have been used to detect the 21\,cm power-spectrum at low redshifts \citep{Chang:2010, Masui:2013, Switzer:2013, Anderson:2018}, many have been turning to interferometers for obtaining the necessary high sensitivities for detecting 21\,cm at higher redshifts. Interferometric experiments seeking to detect 21\,cm fluctuations include CHIME \citep{Bandura:2014}, Tianlai \citep{Chen:2015}, Ooty \citep{Subrahmanya:2017}, HIRAX \citep{Newburgh:2017}, the MWA \citep{Tingay:2013}, LOFAR \citep{VanHaarlem:2013}, the LWA \citep{Ellingson:2009}, and HERA \citep{DeBoer:2017}. Interferometric data sets consist of cross-correlations (visibilities) measured by pairs of antennas (baselines) at various spectral frequencies. Since line-emission at different distance along the Line-of-Sight ($r_\parallel$) is redshifted to different observed frequencies, one can map observed frequencies to co-moving distance $\nu \underset{\sim}{\propto} x_{\parallel}$. For a given visibility, the Fourier dual of frequency is the delay, $\tau$ between signals arriving at each antenna. Thus $\tau \approx 2 \pi Y^{-1} k_{\parallel}$ where $Y$ is a constant. We refer the readers to \citet{Morales:2004} and \citet{Parsons:2012b} for the full expression. Smooth structures, such as foregrounds, reside at delays smaller then light travel time between the two antennas, $\tau_H$; a phenomena known as the ``wedge'' \citep{Datta:2010, Vedantham:2012, Parsons:2012, Morales:2012,Pober:2013}. The fine-scale 21\,cm fluctuations reside at all delays. A natural analysis choice that has been adopted by most Cosmic Dawn fluctuations experiments is to estimate power spectra by applying a discrete Fourier transform (DFT) either on raw interferometric visibilities \citep{Parsons:2012, Parsons:2014, Ali:2015} or on gridded $u$-$v$ data and/or images \citep{ Chapman:2012, Dillon:2013, Dillon:2015, Jacobs:2016, Trott:2016, Barry:2019} and then squaring. In taking an unpadded DFT along a single axis (we consider the $r_\parallel$ axis for example) one replaces the integral in equation~\ref{eq:CONTINUOUSFOURIERTRANFORM} with a discrete sum over $\Ndata$ sampled data points.
\begin{equation}
    \int dr_\parallel e^{-i k_\parallel^n r_\parallel } \to \Delta r_\parallel \sum_{m=0}^{\Ndata-1} e^{- i n k_\parallel^n \Delta r_\parallel},
\end{equation}
where $\Delta r_\parallel$ is the interval between LoS samples and $k_\parallel^m$  is the $n^{th}$ discrete wavenumber, $k_\parallel^n = 2 \pi n (\Ndata \Delta r_\parallel)^{-1}$, $n \in \left\{0, \dots, \Ndata-1\right\}$.
Since foregrounds are confined to the wedge, these techniques can contain/avoid foregrounds by throwing away/downweighting visibility DFT modes with $\tau \lesssim \tau_H$.

Two realities complicate DFT techniques, both of which are related to incomplete sampling. Firstly, data are sampled over a finite bandwidth with a sharp cutoff at the band edges.  Secondly, flagging (excising) of radio frequency interference (RFI) introduces gaps in frequency sampling with additional sharp edges. The DFTs of incompletely sampled foregrounds have (spectral) side-lobes that often greatly exceed the expected amplitude of the 21\,cm signal.

A number of approaches have been adopted to overcome incomplete data coverage. Most address the problem of finite bandwidth by multiplying data by a tapering function that goes to zero at the band-edges \citep{Thyagarajan:2016, Kolopanis:2019}. These multiplicative tapering or apodization filters smoothly filter the components of the signal at the band edges that is affected by sharp finite sampling features.  
While this leads to signal loss, bringing the foregrounds gradually to zero near the band edges compactifies their footprint in the DFT basis. A number of techniques also exist to deal with flagged channels. Per-baseline delay {\tt CLEAN}ing\footnote{This method applies the two-dimensional {\tt CLEAN} algorithm used in radio astronomy imaging \citep{Hogbom:1974} to one spectral dimension.} 
\citep{Parsons:2012} iteratively peels and fits foregrounds on each baseline with a limited number of smooth discrete Fourier modes, interpolating over the channel gaps. Rather than interpolating with DFT modes, {\tt FASTICA} \citep{Chapman:2012} fits smooth independent components at each line-of-sight (LoS) in a data cube, and subtracts them before performing the DFT into bandpower space. 
$\epsilon${\tt ppsilon} \citep{Barry:2019}, similar to {\tt CLEAN}, interpolates over channel gaps with a DFT eigenbasis via the Lomb-Scargle method \citep{Lomb:1976, Scargle:1982}. Unlike {\tt CLEAN}, it also attempts to interpolate the 21\,cm signal by fitting all DFT modes rather than modes within a low delay window.

Any power-spectrum method involves linear filtering, transforming into a power bandpower basis, squaring, and then normalizing squared band-powers with a linear operator can be described in the quadratic estimator (QE) formalism, including several of the already mentioned techniques. For example, while {\tt FASTICA} iteratively determines a foreground subtraction matrix from the data, the application of this subtraction matrix to data can be cast as an \hbox{QE}. \citet{Tegmark:1997} showed that the optimal (information preserving and minimizing error bars) 
quadratic estimator (OQE) for the component of a Gaussian signal~${\bf x}$, that is completely described by discrete bandpowers, $p^\alpha$ is given by a quadratic estimator where (1) the linear filter is the inverse of the data covariance~$\xcovMat^{-1}$, (2)~the transforming and squaring step is performed by the derivative of the total covariance with respect to each $\alpha^{th}$ bandpower~$\dxcovMatAlpha$, and (3)~the normalization matrix is equal to the inverse of the diagonal of the Fisher information matrix~$\text{Diag}\left(\Fmat\right)^{-1}$.

While this recipe is straightforward, several issues complicate its implementation. Perhaps most glaring is the fact that $\xcovMat$ not actually known to much precision. The low-level component from the 21\,cm signal itself is completely unknown while our ability to characterize our instrument \citep{ Pober:2012,  Neben:2015, Neben:2016, Jacobs:2017, Fagnoni:2019} and low frequency foregrounds \citep{Jacobs:2011, Carroll:2016, Line:2017, Zheng:2017, Eastwood:2018} is currently limited to the $\sim 1$\% level.

This has led to attempts at estimating $\xcovMat$ directly from data \citep{Dillon:2015, Ali:2015} and/or modeling it given our understanding of the foregrounds and instrument \citep{Dillon:2013, Shaw:2014, Trott:2016}. Recent investigations have found that data-driven approaches run a high risk of unintentional signal loss (attenuation of the 21\,cm signal) \citep{Switzer:2015, Patil:2016, Cheng:2018} which, if not corrected, led to highly biased results.
Along the same vein, it is unclear how well model driven covariances must accurately represent the underlying data in order to be effective and whether inaccurate model co-variances face similar signal loss issues associated data derived co-variances.

\citet{Liu:2019} point out that attenuation of cosmological modes does not necessarily constitute signal loss as long as we characterize and correct this attenuation downstream. Indeed, standard normalization choices in the literature are explicitly calculated to undo filtering biases. However great care must be exercised. The assumptions under-girding normalization formulas are (as we shall see) easily violated. 

Normalization matrices are also chosen to ``demix'' the smearing between various bandpowers that arise from the non-identity transfer function of our experiment and data-reduction choices.  Effective foreground filters introduce signal loss to foregrounds but not the 21\,cm signal. 
Since filtering can introduce 21\,cm signal loss, it is useful to determine whether and when one can abandon filtering altogether and mitigate all foreground leakage at the demixing normalization step after bandpowers have been formed.

This paper is part one of a two part series. In it, we demonstrate the existence of a simple, fast, and effective foreground filter that is capable of imparting large amounts of good signal loss on arbitrarily sampled spectrally smooth foregrounds. We examine the properties of this filter compare its performance to the traditional approach of band-power estimation with a windowed DFT. In paper two, we will carefully examine the requirements for successfully demixing and reversing signal loss in the normalization step along with the consequences of violating these requirements.

Our filter is based on a simple, analytic model for $\xcovMat$ which captures the essential features of foregrounds: that they are overwhelming bright compared to the signal, that they occupy a continuum of delays up to some maximum, and that we measure them at a finite number of band-limited frequencies. The computation of this covariance matrix can be performed very quickly, using simple closed-form expressions while its analytic simplicity also allows us to study the origins of its efficacy. Because our filter is diagonalized, under certain circumstances, by Discrete Prolate Spheroidal Sequences (DPSS) \citep{Slepian:1978}, we call our method DPSS
Approximate lazY filtEriNg of foregroUnds ({\tt DAYENU})\footnote{In Hebrew, ``day'' translates approximately to ``sufficient'' and ``enu'' means ``to us''. The acronym refers to the fact that our filter is sufficient to us for removing foregrounds for 21\,cm and other intensity mapping datasets.}. While we discuss {\tt DAYENU} in the context of foreground filtering and power-spectrum estimation for 21\,cm cosmology, {\tt DAYENU} can be applied to intensity mapping with other lines (e.g. CII, CO, Ly$\alpha$) where foreground are distinguished from cosmological fluctuations on the basis of spectral smoothness. 

Our paper is organized as follows. In \S~\ref{sec:FORMALISM}, we review the mathematical formalism for QEs. In \S~\ref{sec:DAYENU}, we introduce our simplified inverse covariance weighting scheme, studying its performance on idealized data, its signal loss properties, and its relationship to DFT filtering. In \S~\ref{sec:SIMULATIONS}, we examine {\tt DAYENU}'s performance in foreground filtering and power spectrum estimation with realistic simulations of foregrounds and 21\,cm fluctuations observed by the Hydrogen Epoch of Reionization Array (HERA) \citep{DeBoer:2017}.

\section{Formalism}\label{sec:FORMALISM}
In this section, we set up our notation and review the formalism of QEs and OQEs.
\subsection{Bandpowers}
The data~${\bf x}$ observed in a fluctuation experiment can be decomposed into foregrounds ($\fvec$), noise ($\nvec$), and cosmological fluctuations ($\svec$). 
\begin{equation}
    \xvec = \fvec + \nvec + \svec.
\end{equation}
Since $\fvec$, $\nvec$, and $\svec$ are independent, $\xcovMat = \langle \xvec \xvec^\dagger \rangle - \langle \xvec \rangle \langle \xvec^\dagger \rangle$ can be decomposed into
\begin{equation}
    \xcovMat = \xcovMat_\text{fg} + \Nmat + \Smat,
\end{equation}
where $\Nmat = \langle \nvec \nvec^\dagger \rangle$, $\Smat = \langle \svec \svec^\dagger \rangle - \langle \svec \rangle \langle \svec^\dagger \rangle$, and $\xcovMat_\text{fg} = \langle \fvec \fvec^\dagger \rangle - \langle \fvec \rangle \langle \fvec^\dagger \rangle$.

Bandpowers are usually defined by decomposing $\Smat$ into a set of response matrices
\begin{equation}
    \Smat = \sum_\alpha p^\alpha \dxcovMatAlpha
\end{equation}
While many authors stick with bandpowers that only describe $\Smat$, \citet{Parsons:2014, Ali:2015, Liu:2014a, Liu:2014b} adopt bandpower definitions where $\xcovMat_\text{fg} + \Smat = \sum_\alpha p^\alpha \dxcovMatAlpha$. The decision to define bandpowers for the signal covariance~$\Smat$ alone versus~$\xcovMat_\text{fg} + \Smat$ is an analysis choice with important consequences that we explore in paper II. Since we do not know the 21\,cm signal a-priori, we don't actually know what the correct bandpowers to use are. Instead, we choose a set of response matrices~$\dxcovMatAlphaHat$ that may not actually be correct. A standard choice for $\dxcovMatAlphaHat$ uses our expectation that the 21\,cm signal is homogenous so that the correlation between temperatures at two locations is given by the continuous Fourier transform of the power-spectrum. Authors usually replace this continuous Fourier Transform with a DFT. Thus, many works (e.g. \citep{Dillon:2015, Trott:2016, Barry:2019, Mertens:2020}) choose $\dxcovMatAlphaHat = \dxcovMatAlpha^\text{\bf DFT}$. For a three-dimensional data-cube, each data-point $x_m$ has an associated co-moving position ${\bf r_m}$ so
\begin{equation}
    \left[\dxcovMatAlphaHat^\text{DFT, 3D}\right]_{mn} \propto \sum_{{\bf k} \in V_\alpha}  e^{-  i {\bf k} \cdot ({\bf r}_m - {\bf r}_n )}
\end{equation}
where $V_\alpha$ are fourier-space bins (cylindrical or spherical) and ${\bf k}$ are wave-numbers given by the DFT of a gridded image.

In this work, we focus on per-baseline QEs employed by PAPER and HERA \citep{Parsons:2012, Parsons:2014, Ali:2015} which operate independently on different baselines at different LSTs. These estimators sacrifice a small amount of sensitivity for short baselines \citep{Zhang:2018} and have the advantage of being analytically and computationally simple to work with. For a per-baseline estimator, ${\bf x}$ is the frequency data from a single visibility at a single LST that has potentially been averaged over many identical copies in a redundant baseline group and many different nights at the same LST. We emphasize that this estimator is distinctive from a multi-baseline estimator where the data are ${\bf x}$ consists of all baselines in our data set (e.g. \citealt{Liu:2014a,Liu:2014b}). The DFT bandpowers used in per-baseline estimators are usually just the squared coefficients of a 1D frequency DFT. If the baselines are all sufficiently close together, each spherical $k$-bin is the same as each $k_\parallel$ bin in the LoS DFT. \citet{Parsons:2014},  \citet{Ali:2015}, and in this paper, we focus on LoS DFT bandpowers
\begin{equation}
    \left[  \dxcovMatAlphaHat^\text{\bf DFT} \right]_{mn} \propto e^{-2 \pi i m n /\Ndata}.
\end{equation}

\subsection{Quadratic Estimators}\label{sec:QEFORMALISM}
 In the QE formalism, we denote our $\Nbp$ estimates of bandpowers~$\hat{p}_\alpha$ to be equal to a normalized linear combination pairwise multiplications of data points,
\begin{equation}\label{eq:QE}
\hat{p}_\alpha = \frac{1}{2} \sum_\beta \mathsf{M}_{\alpha\beta} \xvec^\dagger \Ematbeta \xvec - \biasalphahat,
\end{equation}
where $\Ematbeta$ is one of $\Nbp$ different $\Ndata \times \Ndata$ matrices (one for each bandpower) that perform a weighted sum over pairs of data measurements. $\Mmat$ is an $N_\text{b} \times N_\text{b}$ normalization matrix and $\biasalphahat$ is a subtracted estimate of the true bias $\biasalpha$ which includes all covariance contributions not described by bandpowers. 
\begin{equation}
    \biasalpha = \sum_\beta \mathsf{M}_{\alpha\beta} \text{tr}\left[\Ematbeta \left(\xcovMat - \sum_\gamma \dxcovMatGamma \right)\right].
\end{equation}
It is convenient to expand $\Ematalpha$ into a product of filter matrices, $\Rmat$, and a quadratic matrix, $\Qmatalpha$:
\begin{equation}\label{eq:EQE}
\Ematalpha    = \Rmat^\dagger \Qmatalpha \Rmat. 
\end{equation}
Under this expansion, $\Rmat$ describes all filtering applied to data prior to Fourier transforming. For a single visibility, this could be the apodization by a Blackman-Harris window in which case $\mathsf{R}^\text{BH}_{mn} \equiv \delta_{mn}^k T_n^\text{BH}$, where $\boldsymbol{\delta^k}$ is the Kronecker delta matrix and $T_n^\text{BH}$ is the $n^{\mathrm{th}}$ element of a Blackman-Harris window.
Alternatively, for inverse covariance weighting, we might set $\Rmat^\text{OQE} \equiv \xcovMat^{-1}$. $\Qmatalpha$ performs the transformation into the bandpower basis for both data vectors along with binning and squaring. 
A standard example for $\Qmatalpha$ used to estimate DFT bandpowers is the per-baseline delay-transform matrix
\begin{equation}
    \left[\Qmatalpha^{\text{\bf DFT}}\right]_{mn} = e^{-2 \pi i \alpha (m-n) / \Ndata}.
\end{equation}
$\Mmat$ is usually chosen in a way that trades off mixing between band-powers and their error correlations. 
The expectation value of each estimated bandpower, $\hat{p}^\alpha$ is equal to an admixture of true bandpowers
\begin{equation}\label{eq:W}
\left \langle \phat_\alpha \right\rangle = \sum_\beta \Welt_{\alpha \beta} p_\beta + \biasalpha - \biasalphahat
\end{equation}
where 
\begin{equation}
\Wmat = \Mmat \Hmat
\end{equation}
and
\begin{equation}\label{eq:HMAT}
    \mathsf{H}_{\alpha\beta} = \frac{1}{2} \text{tr} \left(\Rmat^\dagger \Qmatalpha \Rmat \dxcovMatBeta \right).
\end{equation} 


\subsection{Optimal Quadratic Estimators}\label{ssec:OQEFORMALISM}
The optimal quadratic estimator that minimizes error bars and preserves all information from the original data is given by \citep{Tegmark:1997,Liu:2011},
\begin{equation}\label{eq:OQE}
    \phat^\alpha_\text{OQE} = \left[\text{Diag}(\Fmat)\right]^{-1}_{\alpha\alpha} \left[  (\xcovMat^{-1} \xvec)^\dagger \dxcovMatAlpha (\xcovMat^{-1} \xvec) \right] - \biasalpha,
\end{equation}
where $\text{Diag}(\Fmat)$ is the diagonal of the Fisher information matrix given by 
\begin{equation}
\mathsf{F}_{\alpha\beta} = \frac{1}{2} \text{tr} \left[ \xcovMat^{-1} \dxcovMatAlpha \xcovMat^{-1} \dxcovMatBeta \right]. 
\end{equation}
If we instead choose, $\Mmat = \Fmat^{-1}$, $\phat_{\text{OQE}}$ also has the desirable property that its window functions are Kronecker deltas so that no mixing between bandpowers occurs. However, fluctuations from the mean, described by the bandpower covariance matrix 
\begin{equation}
    \Sigma_{\alpha\beta} \equiv \langle \hat{p}_\alpha \hat{p}_\beta^* \rangle - \langle \hat{p}_\alpha \rangle \langle \hat{p}_\beta^* \rangle 
\end{equation} are significantly larger and more correlated \citep{Liu:2011}. 

Comparing equation~(\ref{eq:OQE}) with equations~(\ref{eq:QE}) and~(\ref{eq:EQE}), one can plainly see that the OQE is a result of choosing $\Rmat^\text{\bf OQE} = \xcovMat^{-1}$, $\Qmatalpha^\text{\bf OQE}=\dxcovMatAlpha$.

\section{{\tt DAYENU}--A Simple Foreground Filter}\label{sec:DAYENU}

Unfortunately, many of the ingredients in equation~\ref{eq:OQE} including $\xcovMat^{-1}$ weights, $\biasalpha$, and $\Fmat$, require perfect knowledge of $\xcovMat$ which includes thermal noise, the 21\,cm signal, and instrumental effects such as antenna gains. Moreover, our understanding of the radio sky and radio interferometers is limited.  We also don't really know what the correct $\dxcovMatAlpha$ are either -- the focus of paper II. In order to implement an \hbox{OQE}, several authors attempted to estimate $\xcovMat$ directly from the data. \citet{Dillon:2015} obtained $\xcovMatHat$, an estimate of $\xcovMat$ for the frequency-frequency covariance of three-dimensional gridded visibilities by treating all other visibilities in an annulus of fixed $u$ as 
independent samples of the same covariance, ignoring correlations in $u$. \citet{Ali:2015} implemented a per-baseline OQE  $\xcovMatHat$ by computing  the covariance between channels of an individual baseline over time. In that case, because $\xcovMatHat$ is derived from the data itself, there exists significant risk of signal loss \citep{Cheng:2018}. Loss issues led the PAPER team to seek simpler alternatives to $\xcovMat$ estimation. In their most recent analysis, PAPER implemented a per-baseline QE identical to a windowed Fourier transform with $\Rmat = \Rmat^\text{BH}$, $\Mmat = \Imat \equiv \MmatId$, and $\Qmatalpha = \boldsymbol{\mathsf{Q}^{\alpha,\text{DFT}}}$ \citep{Kolopanis:2019}. 

Unfortunately, conservative taper-only filtering choices are of limited utility since they are unable to directly address the sidelobes from incomplete frequency sampling resulting from RFI flags. {\tt CLEAN}ing provides a pre-processing option that can remove a significant fraction of this ringing but has the drawbacks that it is slow and the resulting statistics are difficult to propagate into a final estimate. Furthermore, under realistic flagging conditions, no implementation of 1D {\tt CLEAN} has yet been shown to provide the level of foreground subtraction necessary for a robust 21\,cm detection. Thus, relying on {\tt CLEAN} is a significant risk. A second approach is to model the foreground covariance given our best understanding of the sky's statistics and our radio telescope. Works such as \citet{Shaw:2014} and \citet{Trott:2016} construct detailed models of diffuse and point-source foregrounds and incorporate information on the instrumental primary beam and antenna gains. Modeling approaches are a promising alternative to data-driven covariances that seemingly avoid the associated signal loss risks. However, it is not yet understood what amount of detailed modeling needs to be included in an inverse covariance filter for it to provide sufficient foreground suppression, especially when our knowledge of the instrument and radio sky are so limited.  In this work, we explore a third option; modeling our covariance using as little knowledge of our telescope and foreground statistics as possible ({\tt DAYENU}). 

\subsection{What Makes a Covariance Model Good Enough?}\label{ssec:GOODENOUGH}
Before we construct a simple covariance filter, we should get a sense of what the requirements on an inverse covariance filter are by writing down its action on a data vector. 

If $\Qmatalpha$ performs an untapered Fourier transform, then any foregrounds that are left in our data at this point will be smeared by RFI gaps and the finite bandwidth. Thus, we want the ratio between foregrounds and signal in our inverse covariance-weighted data to be smaller then the level of side-lobes from finite bandwidth and RFI gaps.

To see what requirements this demand puts on our covariance model, we can decompose a hypothetical, non-singular covariance model $\xcovMatHat$ into the sum of eigenvalue-weighted outer-products of its eigenvectors which we divide into a set that are dominated by signal $\{ \uvec_s \}$ and a set that our dominated by foregrounds $\{\uvec_f \}$. 
\begin{equation}
    \xcovMatHat = \sum_{s} \lambda_s \uvec_s \uvec_s^\dagger + \sum_{f} \lambda_f \uvec_f \uvec_f^\dagger.
\end{equation}
The action of $\xcovMatHat^{-1}$ on a data vector $\xvec$ as
\begin{align}
  \zvec \equiv \xcovMatHat^{-1}   \xvec &=   \sum_{s} \frac{1}{\lambda_s} \uvec_s (\uvec_s^\dagger \cdot \xvec) + \sum_{f} \frac{1}{\lambda_f} \uvec_f (\uvec_f^\dagger \cdot \xvec) \nonumber \\
  & = \sum_s \frac{1}{\lambda_s} \uvec_s x_s + \sum_f \frac{1}{\lambda_f} \uvec_f x_f \label{eq:COVFIT}
\end{align}
where $x_s$ are the coefficients of each signal-dominated mode in the data-vector and $x_f$ are the coefficients of each foreground-dominated mode in the data. 
We see in equation~\ref{eq:COVFIT} that all our inverse covariance weighting does is down-weights modes that we have identified as foregrounds in our covariance by $\lambda_f$ and signal by $\lambda_s$. As long as $\lambda_f$ is larger then $\lambda_s$ by the dynamic range between the signal and the foregrounds, then $\zvec$ is dominated by signal. Note that it doesn't actually matter that we get the $\lambda_f$ values right. They just have to be large enough to make the foreground terms much smaller then the signal terms. This is not typically difficult, especially since $\lambda_f$ and $\lambda_s$ square any estimate of the dynamic range between foregrounds and signal so even if an estimate of the dynamic range is low, it is made up for in the squaring. 

We can go one step further and set $\lambda_s = 1$ so that our inverse covariance-weighted vector~$\zvec$ includes signal modes with unity weight and foreground modes that are downweighted by $\lambda_f \gg 1$. As long as we come up with a model covariance whose foreground component is described a relatively small number of orthonormal modes and these modes span the actual foregrounds, the relative amplitudes of the foreground components in our covariance don't actually matter as long as they are large enough to suppress the foregrounds in the data below the signal. While this is a straightforward requirement, it means that regularization factors larger then the signal-foreground dynamic range will spoil foreground subtraction. For example, if $\xcovMatHat$ includes the thermal noise component of a visibility after a short integration, as is the case in \citealt{Dillon:2015, Ali:2015, Trott:2016}, then it may actually prevent sufficient foreground subtraction for a 21\,cm detection even though the covariance is technically more representative of the true data.

To summarize, we have shown that a $\xcovMatHat$ is good enough for 21\,cm power-spectrum estimation in the presence of missing data (RFI gaps and finite, untapered bandwidth) when it upweights all of the principal components of the foregrounds to larger then the dynamic range between foreground and signal modes in the data. The detailed amplitudes of each mode in the actual covariance does not matter as long as the dynamic range is large enough. Covariance models that include thermal noise for short integrations may not include sufficient dynamic range. We can avoid downweighting signal entirely by setting $\lambda_s$ to unity in an estimated covariance by including only foreground modes with large $\lambda_f$ added to an identity matrix.

In the remainder of this section, we will derive a simple covariance matrix that meets these requirements, motivated by the fact that foregrounds are overwhelmingly contained to large wavelength frequency fourier modes over a finite range of delays. The covariance that we do derive will be diagonalized by DPSSs which are a set of vectors whose Fourier coefficients are maximally concentrated to within a finite delay-range. This basis is optimal in the sense that its vectors have maximal dot-products with foregrounds on large frequency scales and minimal dot-products with the 21\,cm signal at fine frequency scales and is an excellent choice for modeling and subtracting band-limited foregrounds in 21\,cm experiments.

\subsection{Defining {\tt DAYENU}}
As a first step towards understanding the necessary modeling fidelity required for effective foreground subtraction we attempt to write a model covariance that makes only the simplest assumptions about the foregrounds on an individual baseline. It has long been appreciated that if we could somehow take a continuous and infinite frequency Fourier transform of a visibility with an achromatic beam, that the power from spectrally flat foregrounds is completely contained to delays with amplitudes less then $\tau \leq \tau_H = b/c$, where $c$ is the speed of light and $b$ is the separation between the two antennas forming the visibility \citep{Datta:2010, Vedantham:2012, Morales:2012, Parsons:2012}. Beam chromaticity and realistic spectral slope and curvature in the foregrounds modify this result but as long as these effects are relatively smooth \citep{ EwallWice:2016b,Thyagarajan:2016,Patra:2018}, they still allow one to define some delay $\tau_w \gtrsim \tau_H$ below which foregrounds are much brighter than any 21\,cm contribution and above which foregrounds are much smaller then both their $\tau=0$ value and 21\,cm fluctuations.

For a particular baseline, we make the simple assumption that the power in each delay is uncorrelated, an assumption that is true for point-source foregrounds but not strictly true for diffuse emission. This is because different delays map to different regions on the sky. \citet{Blake:2002} finds source correlations fall below $\approx 10^{-3}$ on large scales greater then $1^\circ$, thus the different delays for different regions are approximately uncorrelated. Since diffuse emission in different regions of the sky is correlated, diffuse emission in different delays is correlated. In order for delays to be uncorrelated, we must also impose an assumption that the statistics in frequency space are staionary (frequency independent). 

When $\tau \leq \tau_w$ (foreground region), we assume that the variance of each delay is the inverse of a small number~$\epsilon$. For $\tau \geq \tau_w$, we set the variance equal to the channel-width~$\Delta \nu$. 

\begin{equation}\label{eq:LAZYCOVDELAY}
    \ftxcovMatLazyEl(\tau,\tau^\prime)  =  \begin{cases} \epsilon^{-1} \frac{1}{2 \tau_w} \delta^D(\tau-\tau^\prime) & |\tau| \leq \tau_w \\
    \Delta \nu\, \delta^D(\tau-\tau^\prime) & |\tau| > \tau_w.
    \end{cases}
\end{equation}
Here, $\Delta \nu$ is the width of each frequency channel and not necessarily the spacing between different channels. The first piece of equation~\label{eq:LAZYCOVDELAY} represents foregrounds in delay-space while the second piece represents thermal noise.

Suppose we have measurements at $\Ndata$ different arbitrary frequencies. The covariance matrix for these discrete measurements can be obtained by integrating the continuous delay covariance: 
\begin{align}\label{eq:LAZYCOV}
    \xcovMatLazyEl_{mn} & = \int d\tau d\tau^\prime e^{-2 \pi i (\tau \nu_j-\tau^\prime \nu_k)} \ftxcovMatLazyEl(\tau, \tau^\prime) \nonumber \\
    & = \epsilon^{-1} \text{Sinc} \left[2 \pi \tau_w (\nu_m - \nu_n) \right] + \Delta \nu \delta^D(\nu_m - \nu_n) \nonumber \\
    & = \epsilon^{-1} \text{Sinc} \left[2 \pi \tau_w (\nu_m - \nu_n) \right] + \delta^k_{mn},
\end{align}
where $\text{Sinc}[x]\equiv \sin x / x$. In the last line of equation~(\ref{eq:LAZYCOV}), we substitute the Dirac delta-function for a Kronecker delta,\footnote{
This standard normalization for replacing the Dirac delta with the Kronecker delta ensures that $1 = \int d \nu \delta_D =  \Delta \nu \sum_n \delta_{mn}^K / \Delta \nu$.}
$\Delta \nu \delta_D \to \delta^k$. An astute reader might note that we could have just as easily have constructed $\ftxcovMatLazy$ as being diagonal in {\it discrete} delay space instead of continuous delay space and constructed $\xcovMatLazy$ by taking the two-dimensional DFT of $\ftxcovMatLazy$ instead of performing the integrals in equation~\ref{eq:LAZYCOV}. We will justify our choice of a continuous definition in \S~\ref{ssec:DFT} but for now we emphasize that defining $\ftxcovMatLazy$ in continuous delay-space is essential to its efficacy.

In equation~(\ref{eq:LAZYCOV}), we assumed that foregrounds uniformly occupy a finite range of delays between $-\tau_w$ and $\tau_w$. More generally, we can model foregrounds occupying any number of rectangular delay regions (indexed by $\ell$) with half widths of $\tau_w^\ell$ centered at $\tau_c^\ell$ and uniform amplitude $\epsilon_\ell$. 

\begin{align}
    \xcovMatLazyEl_{mn} &= \delta_{mn}^k + \left[\xcovMatLazyFG\right]_{mn}
\end{align}
where
\begin{equation}\label{eq:LAZYCOVGENERAL}
    \left[\xcovMatLazyFG \right]_{mn}= \sum_\ell \frac{1}{\epsilon_\ell}  e^{-2 \pi i \tau_c^\ell (\nu_m-\nu_n) } \text{Sinc}\left[2 \pi \tau_w^\ell  (\nu_m-\nu_n) \right].
\end{equation}
A covariance with multiple delay regions, such as the one in equation~(\ref{eq:LAZYCOVGENERAL}) can be useful for filtering data with super-horizon artifacts including cable reflections \citep{Dillon:2015, EwallWice:2016a, Beardsley:2016}.

We define our lazy {\tt DAYENU} filter to be the inverse of $\xcovMatLazy$, 
\begin{equation}
    \RmatLazy = \left[\xcovMatLazy\right]^{-1}.
\end{equation}
While $\xcovMatLazy$ is Toeplitz, the actual weighting that we apply to visibility data, $\RmatLazy$ is not (Fig.~\ref{fig:EXAMPLEMATRIX}).

\subsection{Without RFI Flags, $\xcovMatLazyEl$ is Diagonalized by Discrete Prolate Spheroidal Sequences.}\label{ssec:DPSS}

The Sinc foreground component to the covariance in equation~(\ref{eq:LAZYCOV}) is diagonalized by a heavily studied set of orthonormal vectors known as discrete prolate spheroidal sequences \citep[DPSSs,][]{Slepian:1978}. 

Letting $\mathcal{W} = \tau_w \Delta \nu$, \citet{Slepian:1978} define a DPSS~$\boldsymbol{u^{(\alpha)}}(\Ndata, \mathcal{W})$ to be one of the countable orthonormal set of vectors solving the eigenvalue problem
\begin{align}
    \sum_{n=0}^{\Ndata-1} \mathsf{L}_{mn}(\Ndata, \mathcal{W}) u_n^{(\alpha)}(\Ndata, \mathcal{W}) &= \lambda_\alpha(\Ndata, \mathcal{W}) u_m^{(\alpha)}(\Ndata, \mathcal{W})
\end{align}
where
\begin{equation}
    \mathsf{L}_{mn}(\Ndata, \mathcal{W}) = \frac{\sin 2 \pi \mathcal{W} (m -n) }{\pi (m-n)}
\end{equation}
Since $\boldsymbol{\mathsf{L}} = 2 \mathcal{W} \xcovMatLazy_\text{FG}$, the DPSSs also diagonalize $\xcovMatLazy_\text{FG}$. Because $\xcovMatLazy$ is the sum of $\xcovMatLazy_\text{FG}$ and an identity term, DPSSs are also the eigenvectors of $\xcovMatLazy$ as we show numerically in Fig.~\ref{fig:EIGENVECTORS}. Let $\left\{h_n \right\}_{\Ndata}$ be the set of all complex sequences of length $\Ndata$. 
\citet{Slepian:1978} show that $\boldsymbol{u^{(0)}}(\Ndata,\mathcal{W})$~the DPSS with the largest eigenvalue~$\lambda_{0}$ is the unit-norm $\Ndata$ sequence that maximizes the quantity
\begin{equation}
    \mu \equiv \frac{\int_{-\mathcal{W}}^{\mathcal{W}} |H(f)|^2 df}{\int_{-1}^{1} |H(f)|^2 df},
\end{equation}
where $H(f)$ is the DFT of $h_n$ centered at $n=(\Ndata-1)/2$. 
\begin{equation}
    H(f) = e^{-i \pi f(\Ndata-1)}\sum_{n=0}^{\Ndata-1}e^{-2 \pi i n f}  h_n.
\end{equation} 
They also show that $\boldsymbol{u^{(1)}}(\Ndata, \mathcal{W})$ is the vector that simultaneously maximizes $\mu$, has unity norm, and is orthogonal to $\boldsymbol{u^{(0)}}(\Ndata, \mathcal{W})$. More generally, $\boldsymbol{u^{(\alpha)}}\left(\Ndata, \mathcal{W}\right)$ is the vector that simultaneously maximizes $\mu$, has unity norm, and is orthogonal to the vectors in the set $\left\{ \boldsymbol{u^{(\alpha^\prime)}}\left(\Ndata, \mathcal{W}\right): \alpha^\prime < \alpha \right\}$.

It follows that DPSSs have the ideal property of maximally concentrating power into a rectangular region of Fourier space with half-bandwidth $\tau_w$. The DPSS with the largest eigenvalue is the unity norm $\Ndata$ length sequence that concentrates maximal power (as quantified by $\mu$) within $\tau_w$. The DPSS with the second largest eigenvalue is the unity norm $\Ndata$-length sequence that maximally concentrates power within $\tau_w$ and is orthogonal to the DPSS with the largest eigenvalue. Ordering DPSSs by their eigenvalues (largest to smallest), the $\alpha^{th}$ DPSS for $\Ndata$ and $\tau_w$ is the length $\Ndata$ unity-norm sequence that maximally concentrates power within $\tau_w$ and is orthogonal to all $\alpha^\prime < \alpha$ DPSSs.   Thus, our foreground covariance is diagonalized by the basis that most efficiently concentrates power within $\tau < \tau_w$. In the absence of channel flags, DPSS vectors are the eigenbasis of $\xcovMatLazy$. As we discussed in \S~\ref{ssec:GOODENOUGH} though this covariance may not include the detailed information on the true values of $\lambda_f$ for each foreground mode on a particular baseline, as long as $\epsilon^{-1}$ is large enough, it will remove the foregrounds to a small enough level that we can measure the 21\,cm signal in the presence of flagging side-lobes.

\citet{Slepian:1978} also show that the first $ \approx 2 \Ndata \mathcal{W}$  eigenvalues of $\boldsymbol{\mathsf{L}}$, $\lambda_\alpha(\Ndata, \mathcal{W})$, are close to unity after which they rapidly drop to zero. When $\Ndata$ is small, the number of non-zero eigenvalues tends to exceed this number but it becomes increasingly accurate as $\Ndata$ increases. Fitting and characterizing foregrounds with DPSS vectors therefor requires $\approx 2 B \tau_w $ components. 

\begin{figure}
    \centering
    \includegraphics[width=0.5\textwidth]{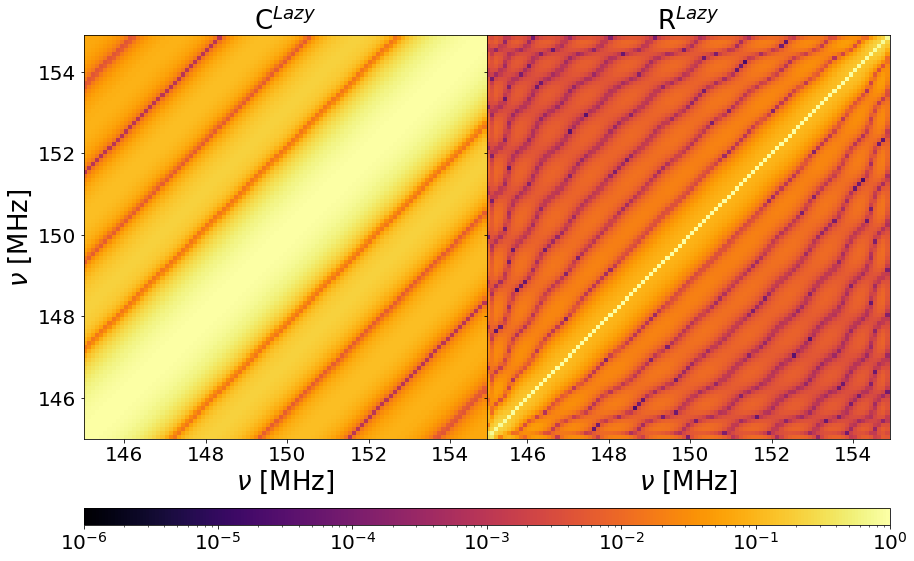}
    \caption{
    Left: An example of $\xcovMatLazy$ for 100 channels, $\Delta \nu = 100$\,kHz, $\epsilon = 10^{-9}$, and $\tau_w = 250$\,ns. $\xcovMatLazy$ is a covariance that is diagonal in the continuous Fourier basis and as a result is Toeplitz.  Right: To obtain a filter matrix, we take the inverse of $\xcovMatLazy$ and obtain $\RmatLazy$. While this inverse is translation invariant in the limit of infinite frequency resolution, it is not for discrete channels. 
    }
    \label{fig:EXAMPLEMATRIX}
\end{figure}

\begin{figure}
    \centering
    \includegraphics[width=.5\textwidth]{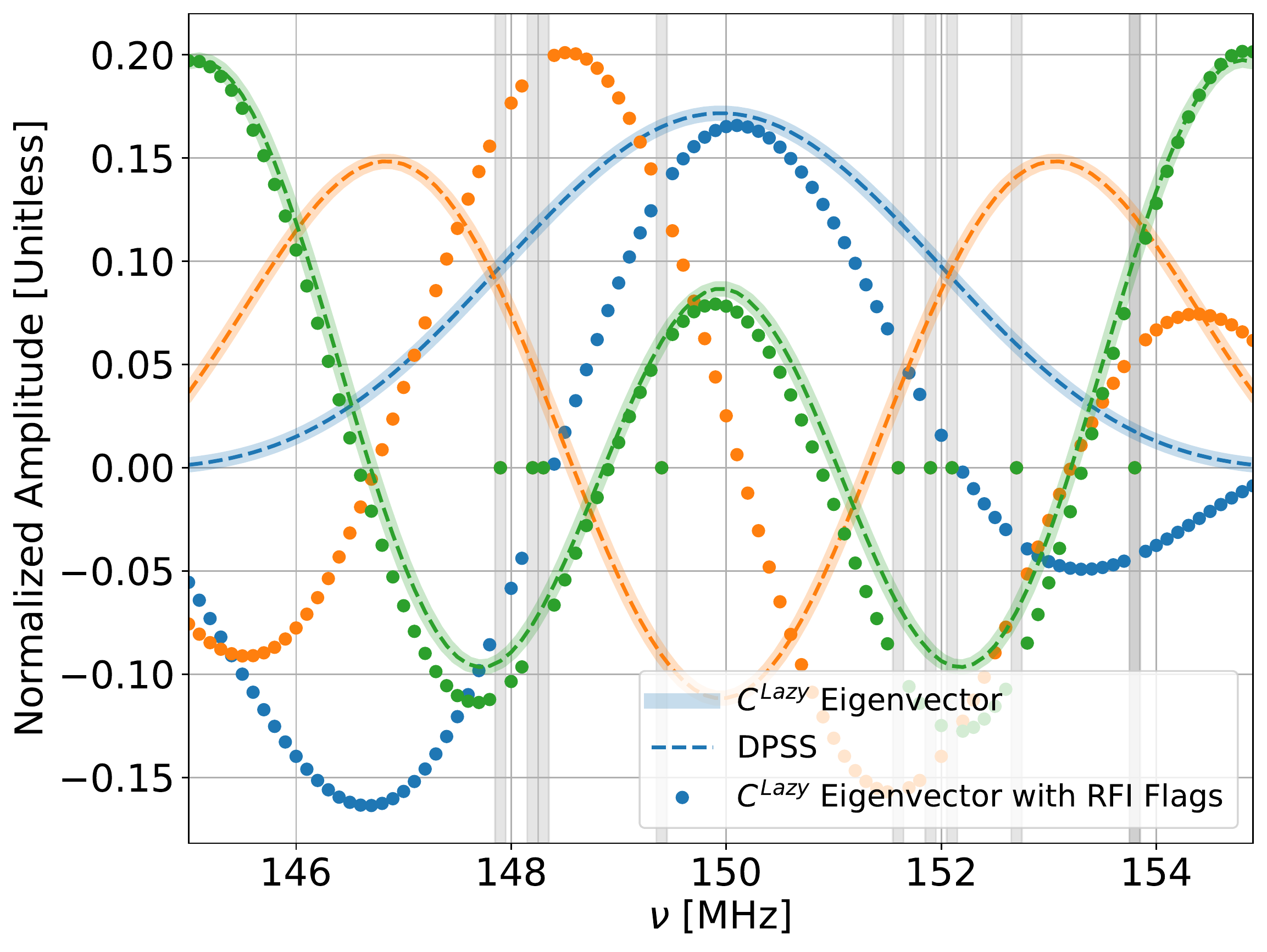}
    \caption{The eigenvectors of the $\xcovMatLazy$ in Fig.~\ref{fig:EXAMPLEMATRIX} with $\Ndata=100$, $B=10$\,MHz, $\tau_w=150$\,ns, and $\epsilon=10^{-9}$ for the zeroth (blue), second (orange), and fourth (green) largest eigenvalues (wide light lines). We compare these eigenvectors to the zeroth (blue), second (orange), and fourth (green) DPSSs  of length $\Ndata=100$, $\tau_w=150$\,ns, over a frequency bandwidth of $B=10$\,MHz (dashed lines). With no flags present $\xcovMatLazy$ is diagonalized by DPSSs. We next set 10 random rows and columns of $\xcovMatLazy$ equal to zero to simulate RFI flags. The resulting eigenvectors (dotted lines) do not correspond to DPSSs.  }
    \label{fig:EIGENVECTORS}
\end{figure}

Under the realistic circumstance that there is missing data (e.g. RFI gaps), the eigenvectors are not equal to DPSSs. In Fig.~\ref{fig:EIGENVECTORS}, we compare the zeroth, second, and fourth numerically determined eigenvectors (ordered by decreasing eigenvalue) of $\xcovMatLazy$ in Fig.~\ref{fig:EXAMPLEMATRIX} to DPSSs with length $\Ndata$, frequency bandwidth $B=10$\,MHz, and delay-space width of $\tau_w = 150$\,ns. To within numerical precision, the DPSSs are identical to numerically computed eigenvectors of $\xcovMatLazy$. We flag ten random channels in $\xcovMatLazy$ by setting the corresponding rows and columns to zero and show the resulting eigenvectors with the zeroth, second, and fourth largest eigenvalues. The eigenvectors of $\xcovMatLazy$ with flagged channels are not merely DPSSs with flagged elements equal to zero. Hence, when we have missing data (RFI gaps), we must set the corresponding rows and columns of $\xcovMatLazy$ to zero and set $\RmatLazy$ equal to the psuedo-inverse of this flagged covariance.

As stated in \S~\ref{ssec:GOODENOUGH}, the effective action of $\RmatLazy$ is to transform our data into a basis close to DPSSs where $\xcovMatLazy$ is diagonal, divide the data by the eigenvalues of $\xcovMatLazy$ in the $\xcovMatLazy$ eigenbasis, and then transform back. The degree to which foreground removal and signal preservation are successful depends on how well isolated foreground and signal components are in the $\xcovMatLazy$ eigenbasis and whether we have included sufficient dynamic range in the $\epsilon^{-1}$ parameter of $\RmatLazy$.

\subsection{A Simple Example.}\label{ssec:SIMPLE}
\begin{figure}
    \centering
    \includegraphics[width=.5\textwidth]{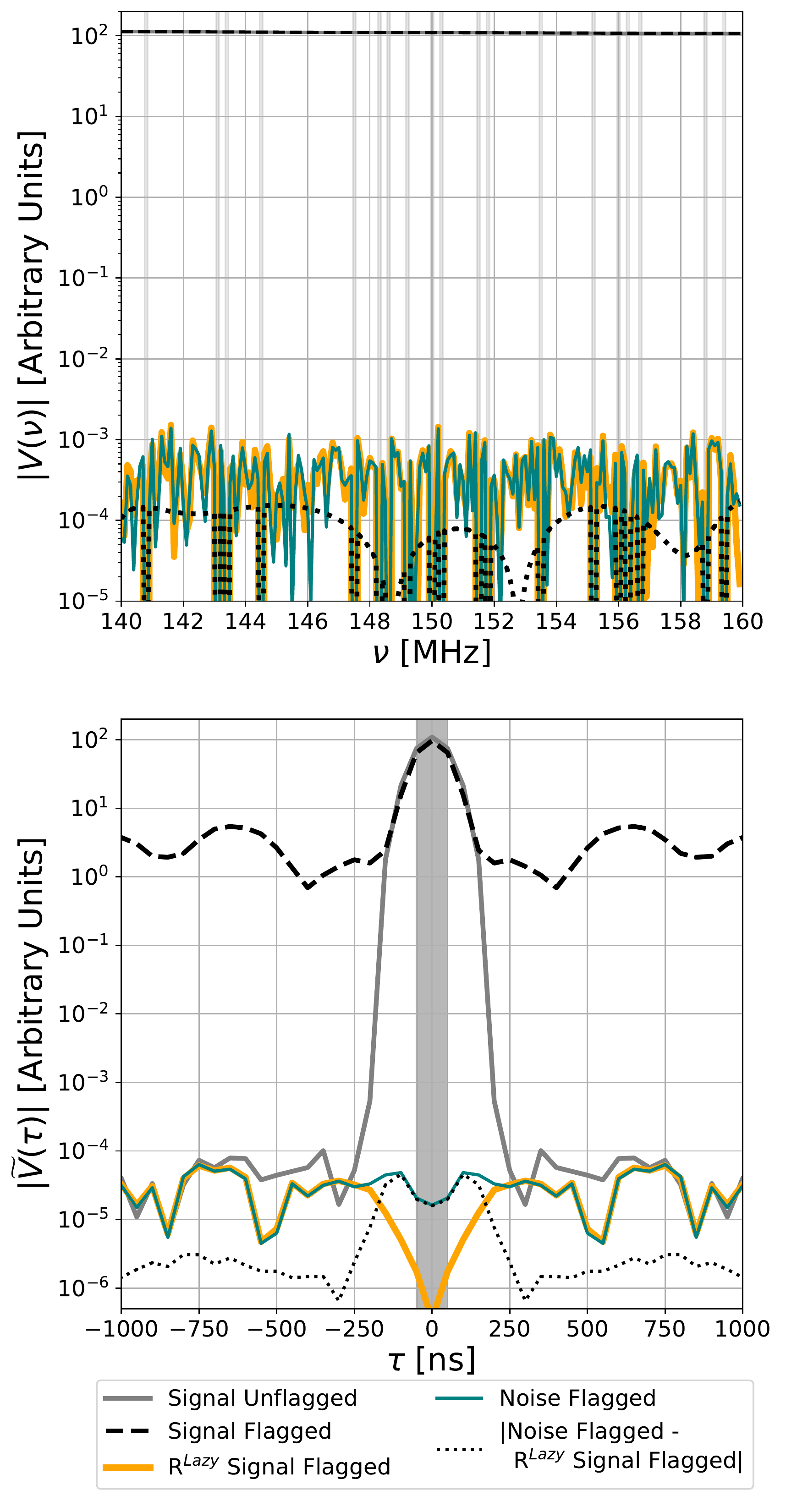}
    \caption{ {\bf Top:} A simulated signal with two hundred channels (noise plus foregrounds) at a single LST drawn from a Gamma distribution with variance consistent with 100 hours of integration, similar to what is necessary for a 21\,cm detection, with (dashed black line) and without (solid grey line) twenty random flags. Flagged channels are shown with vertical grey lines and the corresponding rows and columns in $\xcovMatLazy$ are set to zero before calculating the psuedo-inverse for $\RmatLazy$. Channel-channel fluctuations (thermal noise) are at the $\sim 10^{-5}$ level (orange line). Residuals after applying $\RmatLazy$ with $\tau_w = 50$\,ns, $\epsilon = 10^{-9}$ to the flagged Signal results in the teal curve. The difference between $\RmatLazy$ residuals and the injected noise at the $10$\,\% level (dotted black line). {\bf Bottom:} the same as the top but in the DFT domain (with Blackman-Harris windowing). The filter residual agrees very well with the noise (compare teal and orange in both plots) except for within $100-200$\,ns of the attenuation region (shaded grey rectangle in bottom panel) where some foreground residual is still present. {\tt DAYENU} does not have to down-weight power near the band edges, leading to similar levels of foreground residual across the entire band (dotted black line). Outside of $\sim 200$\,ns, the noise is preserved by the filter at the level of a few percent (compare black dotted and orange lines). 
    }
    \label{fig:AUTOFIRSTLOOK}
\end{figure}

As a first test, we apply it to a realization of a simplistic model autocorrelation for an isotropic sky with temperature $T_\text{sky} = 60\,\mathrm{K}\,(\lambda /1\,\text{m})^{2.55}$,
a chromatic Airy beam from a 14~m-diameter aperture, a receiver temperature of~\hbox{100\,K}, and 200, evenly-spaced frequency channels, of width $\Delta \nu = 100$\,kHz between 140~MHz and~160\,MHz. To simulate RFI flags, we randomly set the power levels in~20 channels to zero. To simulate thermal noise, we assume an integration time of $t_\text{int}=100$\,hr, similar to what is necessary for a robust 21\,cm detection, and set the standard deviation of each channel equal to $A/\sqrt{\Delta \nu t_\text{int}}$ where $A$ is the auto-correlation amplitude \citep{Thompson:2017}. In Fig.~\ref{fig:AUTOFIRSTLOOK}, we show the impact of applying $\left[ \xcovMatLazy \right]^{-1}$ to a single realization of the autocorrelation with $\epsilon=10^{-10}$ and $\tau_w=50$\,ns. After applying our filter, the foregrounds are suppressed by six orders of magnitude and the remaining residual (orange line) is very close to the original noise (green line). Taking the difference between the injected noise and  residuals (dotted grey) we see that in the frequency domain, the filter residuals agree with the injected noise at the $\approx 10\%$ level.

In the bottom panel of Fig.~\ref{fig:AUTOFIRSTLOOK} we inspect our simulation in the delay domain. In the absence of flags, we can use a 7-term Blackman-Harris\footnote{The 7-term Blackman-Harris (see, for example \citealt{Solomon:1993}) includes additional sinusoidal terms beyond the standard 4-term Blackman-Harris found in standard libraries such as {\tt scipy.signal} \citep{Scipy:2020}. While the additional terms increase the width of the central lobe, they substantially lower sidelobes compared to the typical 4-term implementation. We use a 7-term Blackman-Harris taper for all analysis in this paper and refer to it hereon out as simply ``Blackman-Harris''.} taper-filtered Fourier transform to suppress the impact of a finite sampling bandwidth beyond $\approx 250$\,ns (solid grey line). When we set channels containing RFI to zero, these sharp edges spread foregrounds across all DFT modes (black dashed line). We compare the Blackman-Harris Fourier transform of residuals after applying $\RmatLazy$ and the injected noise in delay space. The majority of the $\approx 10\%$ disagreement observed in frequency space is contained within 250\,ns of the edge of our filter (shaded grey region). 

Beyond 250\,ns the injected noise and $\RmatLazy$ residuals agree at the $\approx 10\%$ level. At $\tau \gtrsim 250$\,ns, the leaked foregrounds are subtracted to the level of $10^{-8}$, even with flagging. This is much better than what can be accomplished by an apodized DFT with no flagging. Since apodization functions go to zero at the band edges, they also attenuate the signal. While we applied an apodization before DFTing $\RmatLazy {\bf x}$ to obtain a more direct comparison with with the unflagged model in which no foregrounds were filtered, we technically didn't have. Thus, applying $\RmatLazy$ allows one to circumvent the band-edge signal attenuation that comes with apodization.

In this simplified example, $\RmatLazy$ is highly effective at suppressing foregrounds. However, our simulation made a number of unrealistic assumptions. We assumed an isotropic sky with identical spectral indices. In addition, we assumed that the only chromaticity in our antenna response was sourced by its airy function beam pattern. Ultimately, $\RmatLazy$ and any other inverse covariance filter schemes will only be effective if the foregrounds as viewed by the instrument are spanned by the model covariance's foreground eigenmodes and the model covariance has enough dynamic range to suppress the foreground modes in the data to a level where their flagging side-lobes do not mask the 21\,cm signal power-spectrum. For $\RmatLazy$, this means that it will prevent foreground bleed by the DFT and missing data as long as $\epsilon$ is large enough and $\tau_w$ extends beyond the delays where the foregrounds convolved with the instrument exceed the 21\,cm signal level. From a practical standpoint, this means that $\RmatLazy$ cannot help us detect 21\,cm fluctuations if interal and external antenna reflections as observed for example by \citet{Beardsley:2016, EwallWice:2016, Kern:2019} extend into the delays where interferometers derive most of their sensitivity. On the other hand, if the signal chain chromaticity is contained within some upper $\tau_w$; a design requirement for the Hydrogen Epoch of Reionization Array (HERA) \citep{DeBoer:2017}, then all an analyist needs to do in order to filter foregrounds from their data is to choose a large $\epsilon^{-1}$ and set an appropriate $\tau_w$ in $\RmatLazy$ that extends to the horizon delay $\tau_H$ plus the intrinsic chromaticity of the antenna. Considering HERA as an example; the HERA antenna's chromaticity leaks power above $\approx -50$\,dB at $250$\,ns \citep{EwallWice:2016b, Thyagarajan:2016, Patra:2018}. For HERA, we therefor recommend a $\tau_w$ equal to the wedge plus roughly $250$\,ns.

\subsection{Filtering Efficacy and Signal Attenuation}\label{sec:signalloss}

To be an effective foreground filter, $\RmatLazy$ should attenuate foregrounds while leaving as much of the 21\,cm signal as untouched as possible. If 21\,cm is also attenuated and we do not account for this attenuation in the normalization step we can end up with an unaccounted bias in our measurement: signal loss. Signal loss is not necessarily a bad thing and is in fact desirable if it suppresses foregrounds on otherwise contaminated 21\,cm modes (we would not want our normalization to restore this). In paper II, we will explore when and how good signal loss occurs. In this paper, we focus on the attenuation properties of our simple filter~{\tt DAYENU} with the conservative assumption that we use $\MmatId$ so no correction is made at the normalization step. Under these conditions we treat signal attenuation as significant if its power-spectrum signature exceeds sample variance errors which dominate the most sensitive regions of k-space in upcoming experiments. \citet{Lanman:2019} find that sample variance errors for per-baseline power-spectra are on the order of $20\%$ which places a 10\% constraint on attenuation in the visibility domain.  Spherically averaged power-spectra are expected to be far more sensitive, with $\sim 2\%$ sample-variance errors. This places a constraint of 1\% on visibility attenuation. 

\begin{figure}
    \centering
    \includegraphics[width=.48\textwidth]{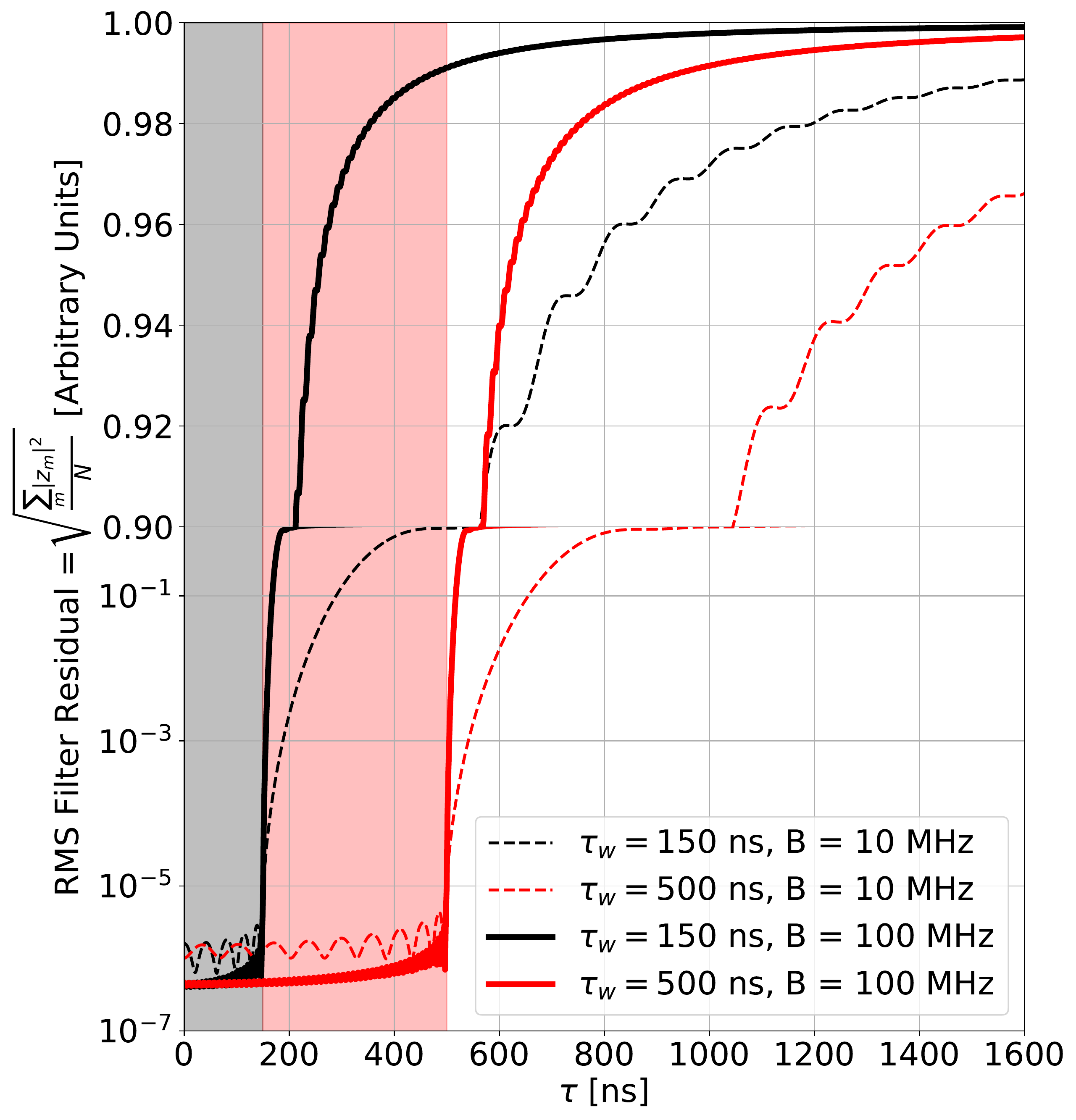}
    \caption{The RMS of residual after applying a $\RmatLazy$ with $\epsilon=10^{-9}$;  $\tau_w = 150$\,ns (black lines) and $\tau_w = 500$\,ns (red lines); and bandwidths of $10$\,MHz (dashed lines) and 100\,MHz (solid lines). Note that the bottom panel has a logarithmic y-scale and the top panel has a linear y-scale. Shaded regions indicate the $\tau_w$ half widths of each filter. Tones within the attenuation region are suppressed between $10^{-7}$ and~$10^{-6}$, more than enough for robust 21\,cm studies. Greater filter bandwidth allows for enhanced overall suppression and reduces attenuation outside of the attenuation region. Attenuation above 10\% is required to bring biases below the level of expected sample variance in per-baseline power spectrum estimates. This occurs for $\tau \gtrsim 300$\,ns beyond the filter edge if a filtering bandwidth of $10$\,MHz is used and only $50$\,ns beyond the filter edge if a bandwidth of~100\,MHz is employed. Spherical power-spectrum estimates will bring variance errors down to $2\%$ in the power-spectrum which translates to a 1\% attenuation requirement in visibility space. Filtering over 100\,MHz brings attenuation below 1\% for $\tau \gtrsim 300$\,ns beyond the filter edge with 100\,MHz of filtering bandwidth. In principal, attenuation can be corrected for at the power spectrum normalization step so these requirements only strictly apply to power-spectrum estimates with identity normalization. 
    }
    \label{fig:INJECTIONBANDWIDTH}
\end{figure}

We investigate the degree that {\tt DAYENU} can suppress modes with different $\tau$ by studying the amplitudes of ${\bf z}^\tau = \RmatLazy {\bf x}^\tau$ where ${\bf x}^\tau$ is a complex sinusoid with delay $\tau$ and amplitude equal to unity sampled every $100$\,kHz. In Fig.~\ref{fig:INJECTIONBANDWIDTH}, we plot the RMS of ${\bf z}^\tau$, $\sqrt{\Ndata^{-1}\sum_m |z_m^\tau|^2}$  vs. $\tau$ for two bandwidths; $10$\,MHz and $100$\,MHz, $\epsilon=10^{-9}$, and two filter widths; $\tau_w=150$\,ns and $\tau_w=500$\,ns.  

Within the attenuation region, we see that input tones are suppressed by a factor of $10^{-7}$ to $10^{-6}$, depending on the bandwidth with larger bandwidths achieving more effective suppression.  When 10\,MHz of bandwidth is used, $\gtrsim 10\%$ signal attenuation occurs within roughly $300$\,ns of the filter edge. Performance improves dramatically if a filtering bandwidth of~100\,MHz is used instead. For 100\,MHz filtering, $\lesssim 10$\% attenuation occurs beyond $50$\,ns of the filter edge and $\lesssim 1$\%  attenuation is reached by 300\,ns beyond the filter edge. Thus, if we conservatively choose to normalize with $\MmatId$ then attenuation beyond $300$\,ns will be smaller then the expected sample variance errors in upcoming experiments. $\MmatId$ is a conservative choice however and we can do better if we choose normalizations that undo these attenuations which we explore in paper II.

\begin{figure}
    \centering
    \includegraphics[width=.5\textwidth]{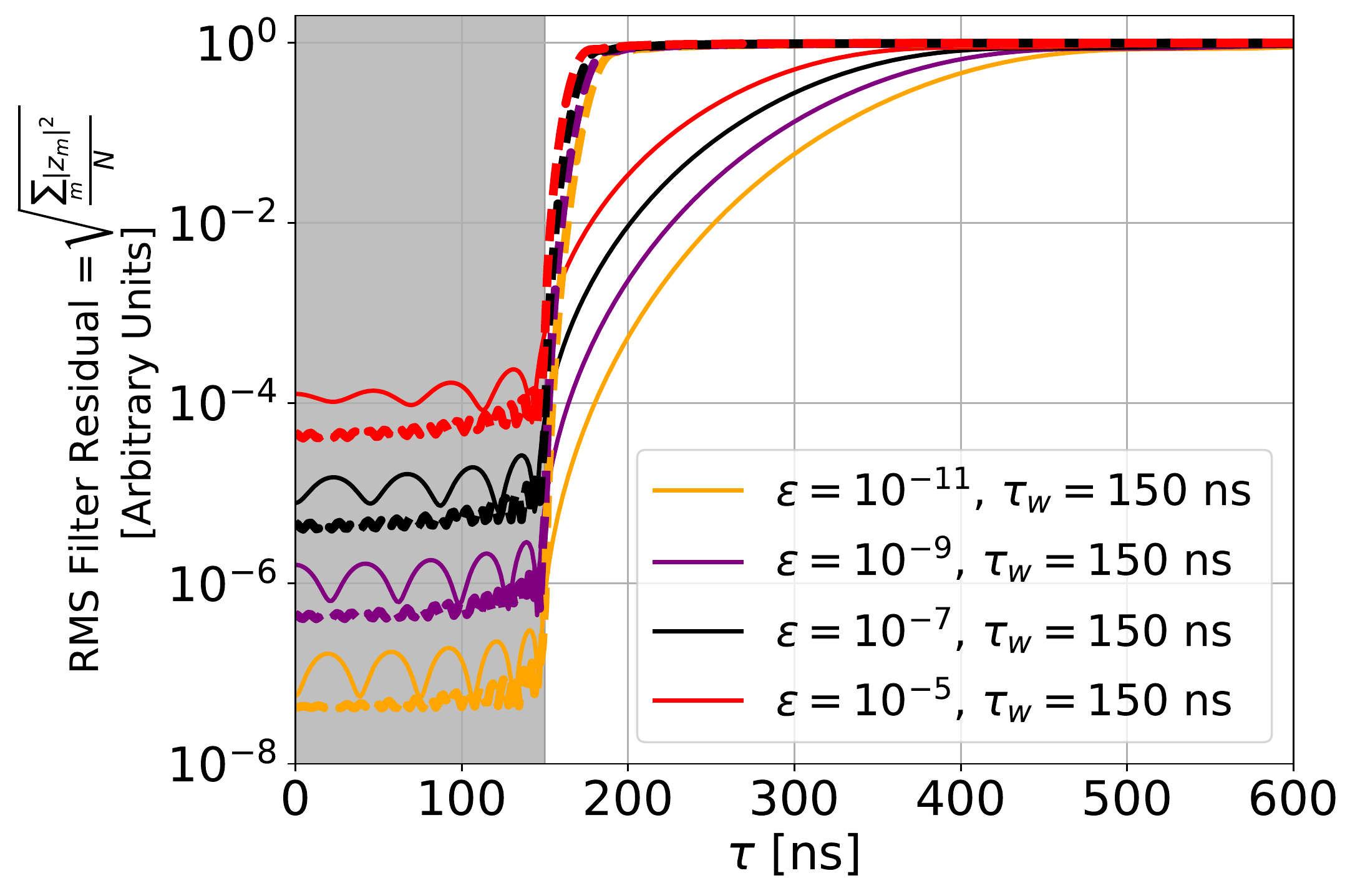}
    \caption{The RMS residual after applying $\RmatLazy$ across 100\,MHz (solid lines) and 10\,MHz (dashed lines) for different values of $\epsilon$. The level of suppression within the filter region is roughly consistent within .25 dex for fixed $\epsilon$ and different bandwidths. }
    \label{fig:VARYEPSILON}
\end{figure}

\begin{figure}
    \centering
    \includegraphics[width=.5\textwidth]{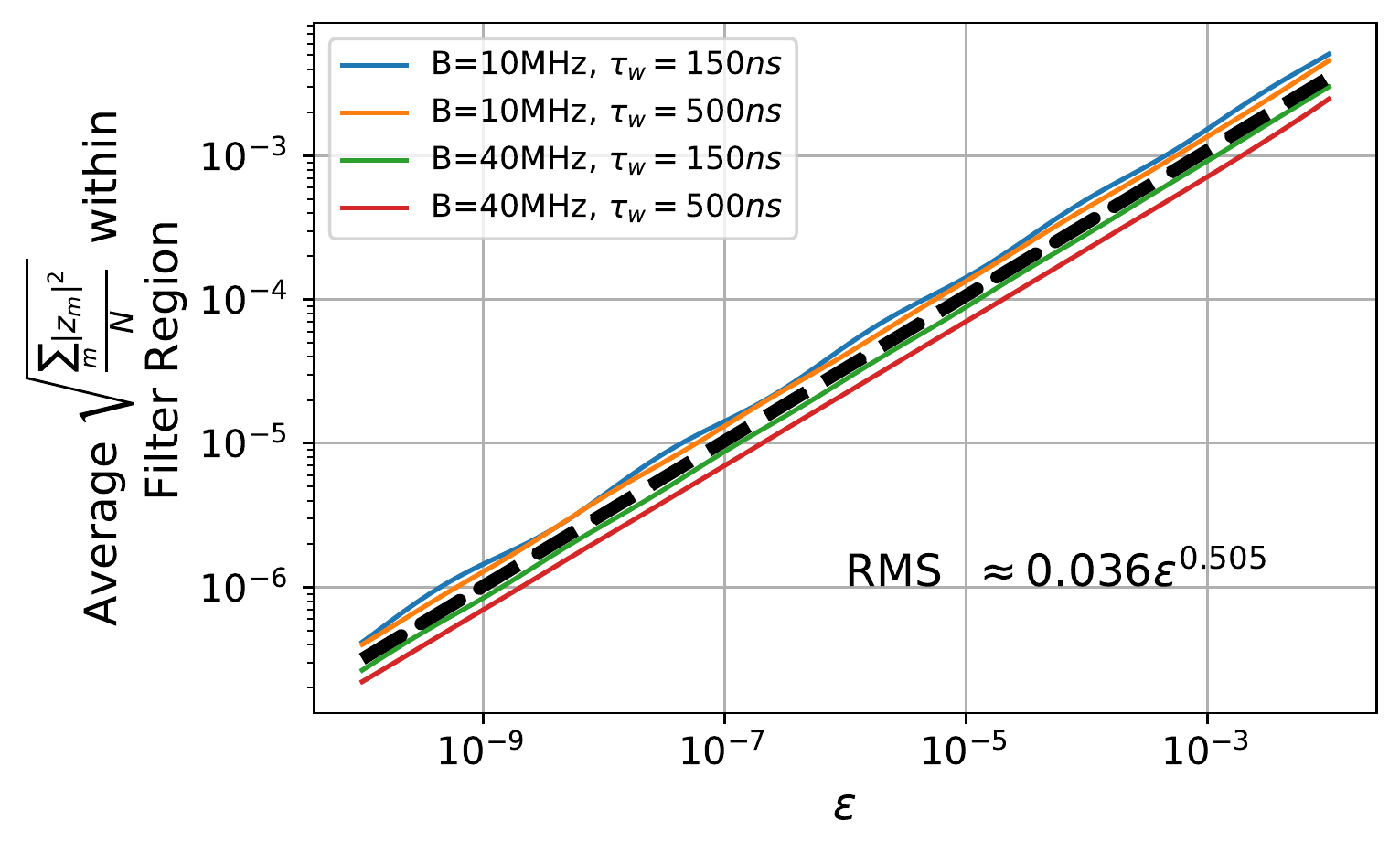}
    \caption{The average suppression of tones within the filter region induced by $\Rmat^\daleth$ for several different filtering bandwidths and $\tau_w$ values (colored lines) as well as a power-law fit to their average (black dashed line). For fixed $\epsilon$, the residual amplitudes inside of the filter region agree within a few dB over a wide range of $\tau_w$ and bandwidths. The RMS residual amplitude goes roughly as the square root of $\epsilon$ (dashed lines). It follows that $\epsilon \lesssim 10^{-8}$ should be used to reduce foreground residuals by a factor of $\approx 10^{-5}$, comfortably below the 21\,cm signal. }
    \label{fig:EPSILONRESID}
\end{figure}

We also inspect how the amplitude of ${\bf z}^\tau$ depends on $\epsilon$ in Fig.~\ref{fig:VARYEPSILON}. We note that the overall level of suppression is consistent (within a few dB) whether we filter across 100\,MHz or 10\,MHz. We compute the average level of suppression of tones over a range of $\tau_w$ and bandwidths as a function of $\epsilon$ in Fig.~\ref{fig:EPSILONRESID}. For a fixed $\epsilon$, the amplitudes of residuals within the filtering region agree within $0.25$\,dex over a wide range of $\tau_w$ and bandwidths. The RMS suppression of Fourier tones within the filtering region follows a power law which we fit to be RMS$\approx 0.1 \epsilon^{0.5}$. It follows that to suppression 21\,cm foregrounds which are $\approx 10^4$ times larger then cosmological fluctuations, we should apply filters with $\epsilon \lesssim 10^{-8}$. Since the foregrounds in the EoR window will be suppressed by flagging side-lobes, it is possible that one could get away with $\epsilon$ one-to-two orders of magnitude larger depending on the severity of flagging.

\subsection{{\tt DAYENU} and the DFT Basis}\label{ssec:DFT}

\begin{figure}
    \centering
    \includegraphics[width=.5\textwidth]{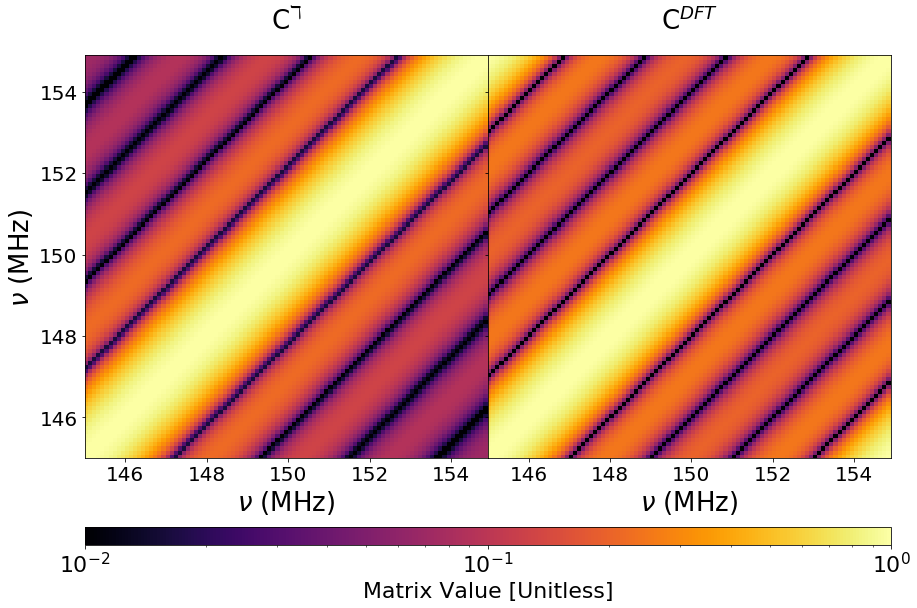}
    \caption{Left: $\xcovMatLazy$ with $\tau_w = 250$\,ns and $\epsilon = 10^{-9}$ where each element is obtained using a continuous Fourier transform (equation~\ref{eq:LAZYCOV}).
    Right: $\xcovMat^\text{DFT}$, the 2D DFT of which is diagonal. The two matrices differ through the presence of wrap-around, $\xcovMat^\text{DFT}$ is equal to an infinite sum of copies of $\xcovMatLazy$ translated by integer intervals of $B$ along the diagonal (equation~\ref{eq:DFTCOVFREQ}). The real-life absence of correlations between opposite band-edges in our foregrounds, which is demanded by DFT modes, is what causes $\xcovMat^\text{DFT}$ to perform poorly relative to $\xcovMatLazy$.
    }
    \label{fig:DFTLAZYCOMPARISON}
\end{figure}

To derive $\xcovMatLazy$ (equation \ref{eq:LAZYCOV}), we wrote down discrete elements of our frequency covariance matrix by taking the continuous Fourier transform of a covariance that was diagonal in continuous delay space. On the other hand, many power spectrum estimators (e.g., \citealt{ Parsons:2012, Dillon:2013, Trott:2016, Barry:2019}) estimate band-powers in DFT space.  This difference in approach immediately raises the question, why not derive $\Rmat$ from a covariance matrix that is diagonal in DFT space rather than the continuous space that we chose? After all, if we could just write down $\Rmat$ as diagonal in DFT space, could we just divide the DFT of our data-set by the diagonal DFT of $\Rmat$,$\boldsymbol{\widetilde{\mathsf{R}}}$, and save computational steps? The short answer is that an $\Rmat$ that is diagonal in DFT space only includes information on foreground modes with delays equal to $m/B, m \in \{-\Ndata/2, \dots \Ndata/2 - 1\}$ and as a result is incapable of properly suppressing foregrounds at intermediate delays.  In order to see this effect, we write $\xcovMat^\text{DFT}$ as the discrete Fourier transform of a covariance that is diagonal in DFT space, 
\begin{equation}\label{eq:DFTCOVDELAY}
    \ftxcovMatDFTEl_{rs} = \begin{cases} \epsilon^{-1} \frac{1}{2 \tau_w B} \delta^k_{rs} & \left|\frac{r}{B}\right| \leq \tau_w \\ 
    \delta^k_{rs} & \left|\frac{r}{B}\right| > \tau_w \end{cases}.
\end{equation}
We then transform 
 $\boldsymbol{\widetilde{\mathsf{C}}}^\text{DFT}$ into discrete frequency space by performing a 2D DFT. 
\begin{align}\label{eq:DFTCOVFREQ}
    \xcovMatDFTEl_{mn} & = \delta_{mn}^k + \frac{\epsilon^{-1}}{2 \tau_w B} \sum_{\substack{|r| \leq \tau_w B\\ | s | \leq \tau_w B}}e^{-2 \pi i (rm - sn) / \Ndata} \delta_{rs}^k\nonumber \\
    & = \delta^k_{mn} + \frac{ \epsilon^{-1} }{2 \tau_w B} \sum_{|r|\leq \tau_w B} e^{-2 \pi i r(m-n)/\Ndata} \nonumber \\
    &= \delta^k_{mn} + \epsilon^{-1} \sum_{s=-\infty}^{\infty} \text{Sinc} \left[2 \pi \tau_w \left(B \frac{m-n}{\Ndata} - s B \right)\right]
\end{align}
where we used the Poisson summation formula (e.g., \citealt{Epstein:2007}) to go from the second and third lines in equation~\ref{eq:DFTCOVFREQ}. We see that the foreground component of $\xcovMatDFT$ is essentially an infinite sum of copies of the foreground component of $\xcovMatLazy$ translated along the diagonal by integer multiples of $B$. 
This can also be seen by visual inspection in Fig.~\ref{fig:DFTLAZYCOMPARISON} where we plot $\xcovMatLazy$ next to $\xcovMatDFT$. The wrap-around arises from the fact that our covariance elements are exclusively comprised of tones that are periodic over the interval $B$. 

By definition, $\xcovMatDFT$ is diagonalized by the DFT. Thus, when we weight by its inverse, it will only down-weight modes with $\tau = m B^{-1} \leq \tau_w$; harmonic or on-grid DFT tones. Visibilities include a continuum of delays and only a fraction of their power is accounted for by harmonic tones within the wedge. Thus, $\Rmat^\text{DFT} \equiv \left[\xcovMatDFT\right]^{-1}$ is incapable of removing the bulk of foreground power, especially power in the sinc-sidelobes of the aharmonic tones. These side-lobes remain at high delays and prohibit a 21\,cm measurement.

\begin{figure*}
    \centering
    \includegraphics[width=\textwidth]{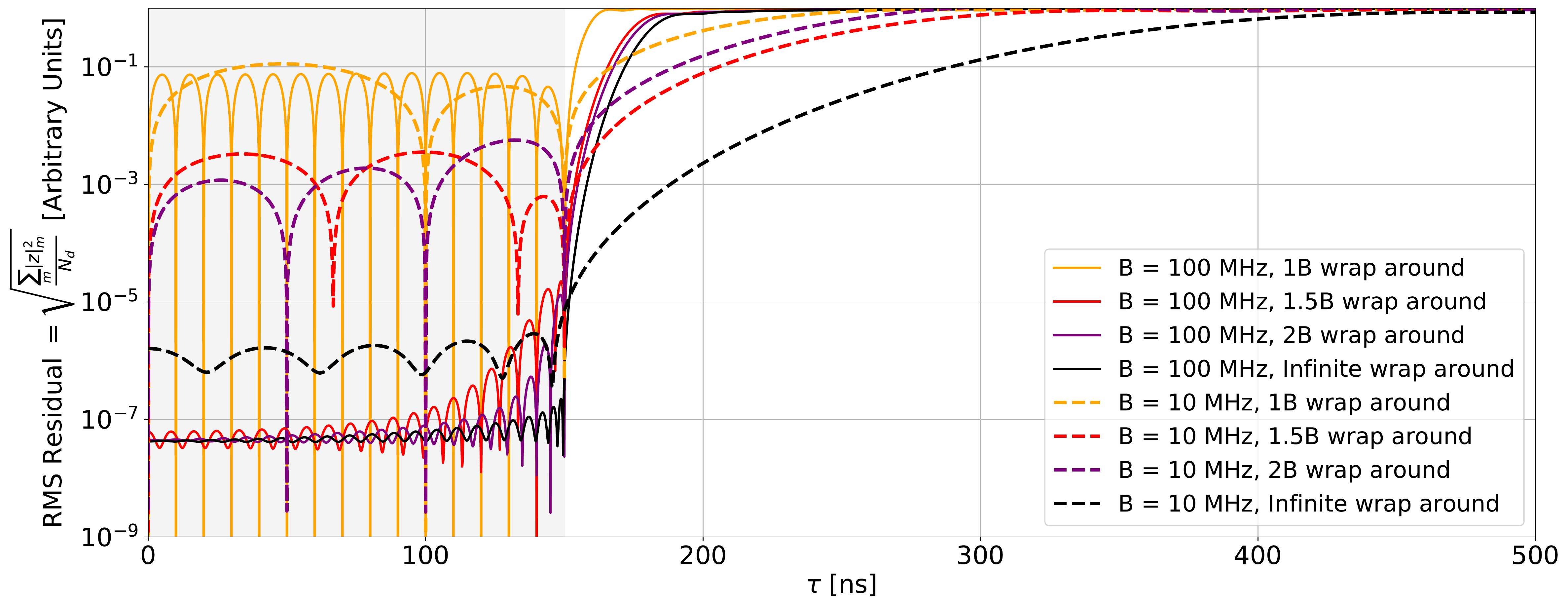}
    \caption{The RMS residual $\RmatLazy$ applied of tones with delay $\tau$. We filter $\tau \lesssim \tau_w \approx 125$\,ns over 10\,MHz (dashed lines) and 100\,MHz (solid lines) with the covariance matrix peridodicity (the coefficient next to `m' in equation~\ref{eq:DFTCOVFREQ}) set to be 1B (grey lines), 1.5B (red lines), 2B (purple lines) and infinite (black lines). 
    Enforcing periodicity on the covariance matrix is equivalent to restricting its Fourier modes to be harmonics of its wrap-around period. As a result, the covariance matrix is only able to effectively filter these harmonics. For example, when we set periodicity to $10$\,MHz, our filter only effectively removes the $1/(10\text{\,MHz}) = 100$\,ns tone (dashed black line). When the periodicity is set to $20$\,MHz, we can remove the 50\,ns, 100\,ns, and 150\,ns tones. When we use $100$\,MHz bandwidth, tones are spaced by $10$\,ns. When we set the periodicity to $200$\,MHz, the spacing between tones drops to $\approx 5$\,ns but all tones within the attenuation region are effectively removed due to the finite width of suppression about each tone. The fact that the DFT diagnalized filtering matrix approximately converges to {\tt DAYENU} at $\gtrsim 2B$ wrap-around indicates that $\sim 4 \tau_w B$ modes must be fit in order to achieve similar performance.  
    This can be understood as an approximate manifestation of Nyquist's theorem since we are attempting to describe frequency-limited foregrounds with infinite but highly concentrated support in delay-space. Representing such a signal requires at least $\gtrsim 1/2B$ sampling. 
    }
    \label{fig:INJECTIONWRAP}
\end{figure*}

Figure~\ref{fig:INJECTIONWRAP} illustrates the limitations of $\xcovMat^\text{DFT}$, where we show the same quantities as in Fig.~\ref{fig:INJECTIONBANDWIDTH} but now include the performance of $\Rmat^\text{DFT}$.  
We study the impact of progressively adding in-between-modes back into $\xcovMat^\text{DFT}$ by increasing the wrap-around interval in equation~\ref{eq:DFTCOVFREQ}. For example, increasing the wrap-around from $B$ to $2B$, adds additional modes that are periodic over a bandwidth of $2B$ but are not periodic over $B$. The orange lines in Fig.~\ref{fig:INJECTIONWRAP} show the residual amplitudes leftover after applying $\Rmat^\text{DFT}$ to complex sinusoides with various delays, $\tau$. Unlike $\RmatLazy$, gaps are present, $\Rmat^\text{DFT}$'s filter coverage and truely effective filtering only occurs at $\tau = m/B, m \in \mathbb{Z}$. Between $B^{-1}$ harmonics, filtering only decreases the foreground amplitude by a factor of $\sim 10^{-1}$. 

As we increase period of the wrap-around in equation~\ref{eq:DFTCOVFREQ}, the harmonic filter tones move closer together and eventually merge. Because larger bandwidths have greater Fourier resolution, increasing the DFT wrap-around to 2B over 100\,MHz actually attains similar performance for the completely continuous case though {\tt DAYENU} still subtracts foregrounds to roughly $\approx 10^{-2} \times$ the level of DFT modes at the filter edge. This indicates that if we did want to use DFT modes to model our foregrounds and subract them, we need on the order of $\gtrsim 2\times$ as many modes. Since $\xcovMatDFT$ converges to {\tt DAYENU} as the wrap interval approaches $\gtrsim 2B$, roughly $\gtrsim 4 \tau_w B$ DFT modes are necessary to model foregrounds at a level similar to $\approx 2 \tau_w B$ DPSS vectors. As we mentioned in \S~\ref{ssec:DPSS}, for large $\Ndata$, the number of DPSS modes with non-zero eigenvalues in $\xcovMatLazy$ is approximately $2 B \tau_w$.

 If the DPSS modes are precomputed and the number of DPSS modes being fit is much less then the number of frequency channels, then finding the fit coefficients for a single
 flagging pattern and set of fitted modes is dominated by calculating $\Amat^\dagger \boldsymbol{\mathsf{w}} \Amat$ where $\Amat$ is the $\Ndata \times N_\text{mode}$ design matrix where each row is one of the $N_\text{mode}$ DPSS vectors that we are fitting. This matrix multiplication requires $\sim \mathcal{O}(\Ndata N_\text{mode}^2)$ operations. Since typically twice as many DFT modes are required then DPSS modes, DPSS fitting with pre-computed modes reduces computational operations by a factor of four.

In summary, filtering with a covariance that is diagonal in the discrete Fourier basis will perform very poorly in foreground subtraction because it only contains the subset of foreground modes that are harmonics of $B^{-1}$. In defining $\xcovMatLazy$, we instead allow foregrounds to include any continuous delay within the wedge and use numerical matrix inversion determine and downweight a discrete set of principal components. 

\subsection{Pre-Truncation Filtering}\label{ssec:TRUNCATION}
It is clear from Fig.~\ref{fig:INJECTIONBANDWIDTH} that the larger the bandwidth we filter over, the smaller the unwanted signal attenuation outside of $\tau_w$. This motivates the use of $\sim 100$\,MHz bandwidths for filtering. The power spectrum is usually approximated over bandwidths of $\lesssim 10$\,MHz in order to ensure roughly stationary statistics for the evolving 21\,cm signal.

These two ends can simultaneously be achieved by applying $\RmatLazy$ over a $\approx 100$\,MHz band, truncating, and then estimating the power spectrum from a DFT over a smaller sub-band. Under this scheme, $\RmatLazy$ is a non-square $\Ndata \times \Ndata^\text{F}$ matrix, where $\Ndata^\text{F}$ is the number of channels to be filtered over and $\Ndata^\text{F} \geq \Ndata$. To obtain a truncated $\RmatLazy$, all we have to do is zero out the rows of $\RmatLazy$ corresponding to channels that we do not want to include in the application of $\Qmatalpha$. 

\begin{figure}
    \centering
    \includegraphics[width=.5\textwidth]{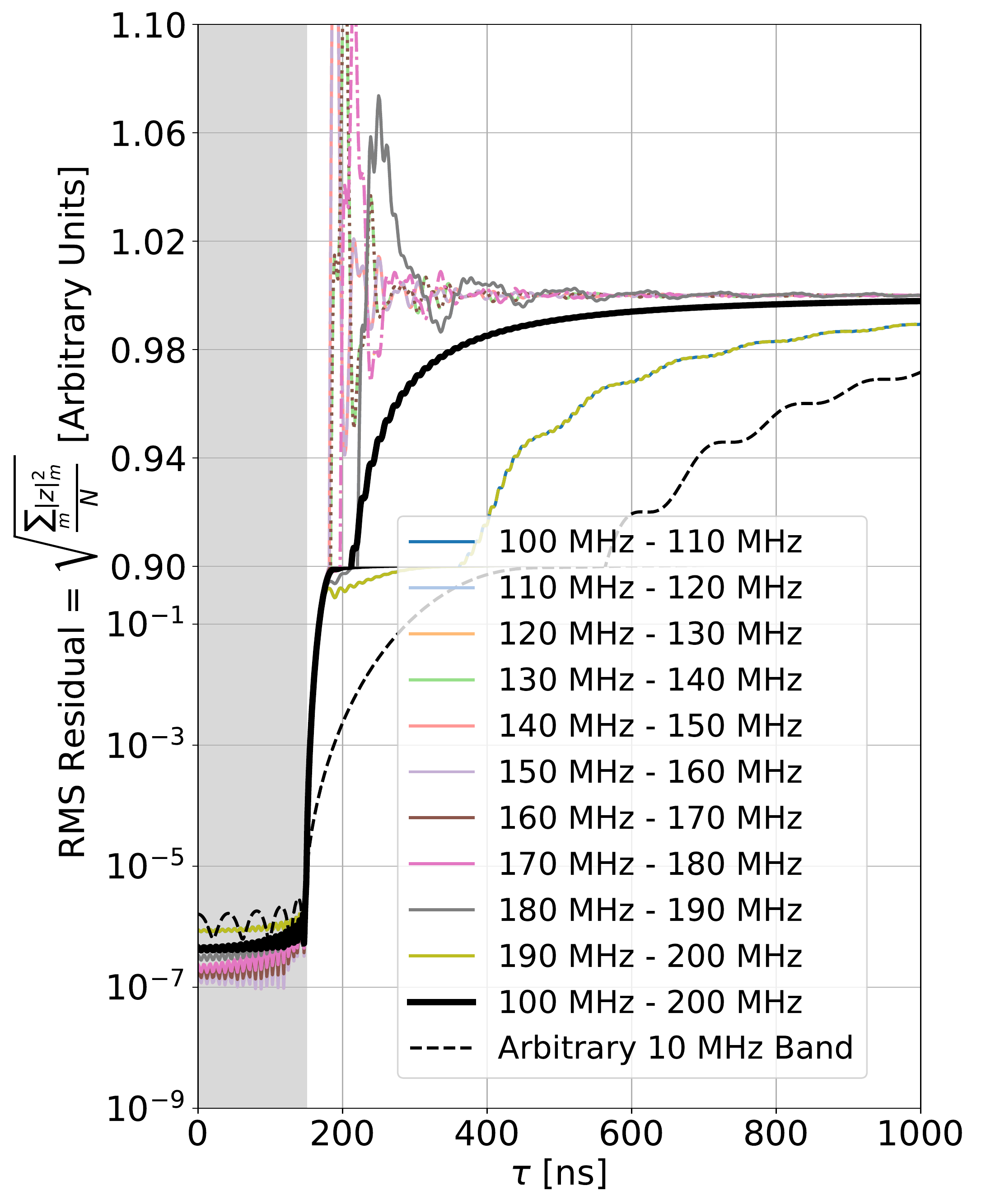}
    \caption{
    The RMS residual of truncated $10$\, MHz subbands of tones after $\RmatLazy$ is applied to the full 100\,MHz band. The degree of signal attenuation is significantly improved over the case where the filter is applied directly to each $10$\,MHz subband after truncation (black dashed line). With the exception of the two outer subbands, signal attenuation is below $1\%$ by $\gtrsim 200$\,ns beyond the filter edge. The edge bands have $\lesssim 10$\,\% signal attenuation within 250\,ns of the filter edge. Bringing attenuation below $\lesssim 1\%$ brings it within the expected sample variance error bars of spherically average power spectra. Bringing this attenuation below $10$\% brings it below the expected sample variance of per-baseline power spectra \citep{Lanman:2019}}
    \label{fig:INJECTIONTRUNCATION}
\end{figure}

Fig.~\ref{fig:INJECTIONTRUNCATION} examines signal attenuation as a function $\tau$ over ten different $10$\,MHz sub-bands where truncation to $10$\,MHz is performed after the application of $\RmatLazy$. In each sub-band, signal attenuation is dramatically reduced compared to  filtering over the 10\,MHz band alone. With the exception of the edge bands (100-110\,MHz and 190-200\,MHz), $\lesssim 1$\,\% signal attenuation is achieved by 250\,ns beyond the filter edge. 
In the outer 10\,MHz bands, 10\% loss is still achieved by 150-200\,ns off the filter edge. In light of these results, we recommend sub-band power-spectrum estimates be obtained from data on which {\tt DAYENU} is applied over as wide a band as possible and then truncated. 

\subsection{Flagged Channels}\label{ssec:FLAGGING}

\begin{figure}
    \centering
    \includegraphics[width=.5\textwidth]{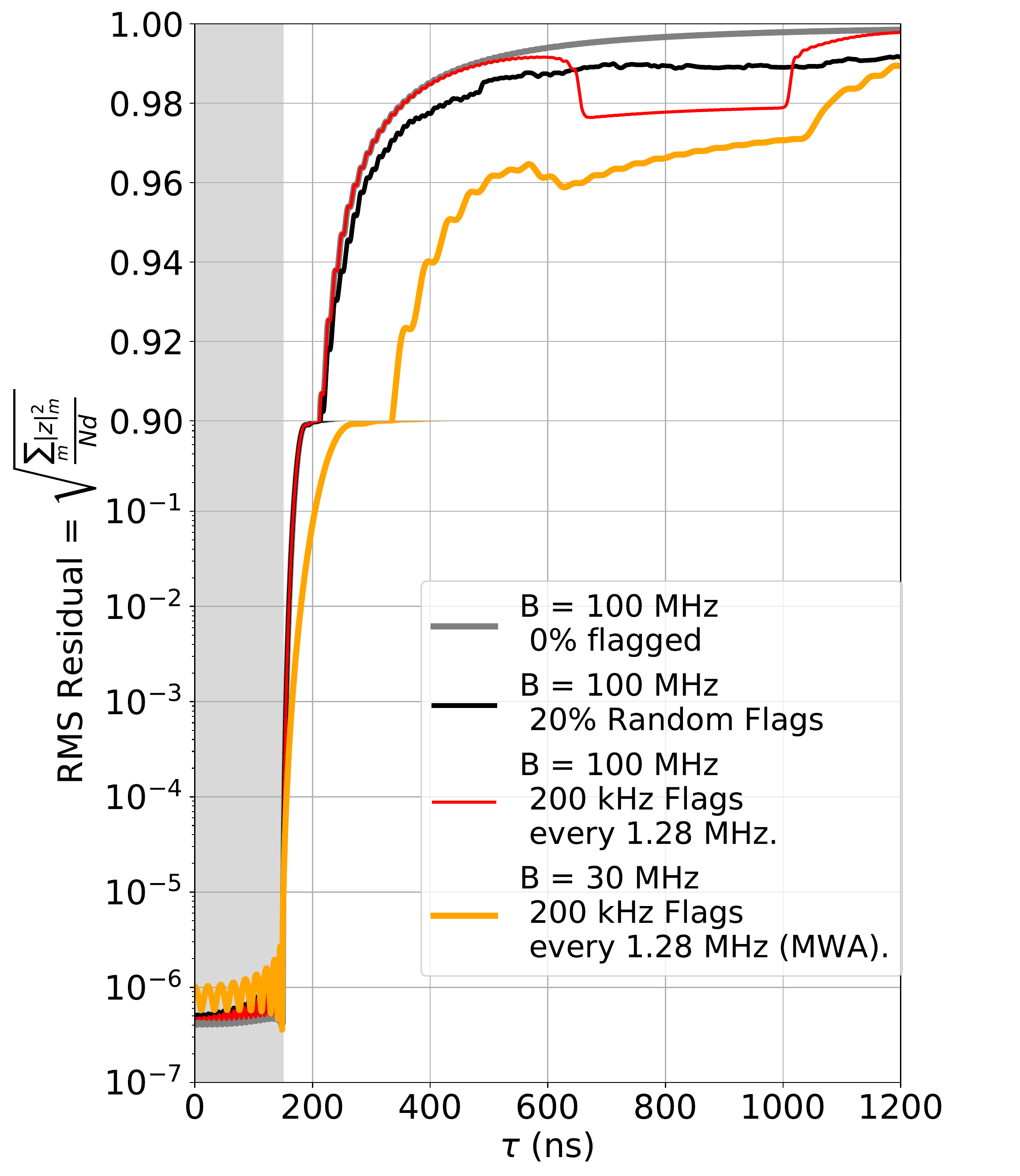}
    \caption{RMS residuals of $\RmatLazy$ for tones filtered with various flagging patterns in data sampled every 100\,kHz. We compare no flags over 100\,MHz (black line) 200 randomly flagged channels (grey line) and 200\,kHz of flagging every 1.28\,MHz (red line) -- similar to what is typically performed on the MWA. Since the MWA only observes 30\,MHz simultaneously, we also show 200\,kHz flags every 1.28\,MHz (gold line). Random flagging increases attenuation by a percent or so. MWA-like flagging results in $\approx 2\%$ attenuation over most delays.}
    \label{fig:FLAGGINGEXAMPLE}
\end{figure}

In real life, some fraction of interferometric channels are contaminated by RFI and must be discarded. Thus, it is necessary for {\tt DAYENU} to work robustly on data that is not evenly sampled. We investigate the impact of RFI flagging by inspecting RMS residuals from applying the psuedo-inverse of $\xcovMatLazy$ where rows and columns corresponding to flagged channels are set to zero. We explore two different scenarios over 100\,MHz of bandwidth. One in which twenty percent of channels are flagged randomly and one in which 200\,kHz flags are applied every 1.28\,MHz; similar to what must be performed on the MWA \citep{Dillon:2015, EwallWice:2016a,Beardsley:2016,Barry:2019} (Fig.~\ref{fig:FLAGGINGEXAMPLE}). Since the MWA records $\approx 30$\,MHz simultaneously, we also show the RMS residual of $\RmatLazy$ with 200\,kHz flags every 1.28\,MHz over 30\,MHz.

WIth 200 100\,kHz channels flagged randomly over 100\,MHz, we find that attenuation beyond the filter width increases by approximately 1\% out to large delays. The presence of periodic flags results in the flagging attenuation being concentrated in a concentrated region centered  $\approx 781$\,ns, the delay of the 1.28\,MHz flag periodicity. Outside of this region, the attenuation is negligible but within this region it exceeds 2\%, in excess of the average 1\% induced by randomized flagging.  

\subsection{{\tt DAYENUREST}}\label{ssec:DAYENUREST}

By subtracting foregrounds with a matrix multiplication, {\tt DAYENU} accomplishes one of the primary objectives of the iterative {\tt CLEAN} filter \citep{Parsons:2012}. ${\bf z}=\RmatLazy {\bf x}$ is equivalent to the residual after {\tt CLEAN} is applied. The second goal of {\tt CLEAN} is to smoothly interpolate (restore) the subtracted foregrounds by adding back their {\tt CLEAN} components; interpolating the foregrounds over flagged channel gaps with DFT modes. We can isolate the foregrounds subtracted by $\RmatLazy$ with the matrix operation $(\Imat - \RmatLazy)$ and fit them to $N_\text{DPSS}$ DPSS modes. 
DPSS vectors are eigenvectors of the foreground component of $\xcovMatLazy$ so we can approximate our foregrounds with the DPSS vectors with eigenvalues above some small number relative to the largest eigenvalues. We choose a cutoff of $10^{-12}$ the largest eigenvalue which ensures that foreground modes are subtracted to a level of $\lesssim 10^{-6}$. 

Fitting and interpolating with our $N_\text{DPSS}$ modes can be achieved applying the linear least squares solution matrix to $(\Imat-\RmatLazy)$. 
\begin{equation}
    \boldsymbol{\mathcal{A}} = \Amat \left[ \Amat^\mathsf{T} \wmat \Amat \right]^{-1} \Amat^\mathsf{T} \wmat
\end{equation}
where $\Amat$ is an $\Ndata \times N_\text{DPSS}$ matrix 

\begin{equation}
    \mathsf{A}_{m\alpha} = u_m^{(\alpha)}(\Ndata, \tau_w)
\end{equation}
where $u_m^{(\alpha)}(\Ndata, \tau_w)$ is the $m^{th}$ element of the $\alpha^{th}$ DPSS vector of length $\Ndata$ that diagnalizes the $\Ndata \times \Ndata$ matrix $\mathsf{S}_{mn}(\Ndata, \tau_w) = (2 \tau_w \Delta \nu )\text{Sinc}\left[2 \pi \tau_w (\nu_m - \nu_n)\right]$ and $\wmat$ is a diagonal matrix set to unity at unflagged channels and zero at flagged channels. 
Applying $\boldsymbol{\mathcal{A}}$ to $(\Imat - \RmatLazy)$ provides us with DPSS interpolated {\tt CLEAN} components. Adding these {\tt CLEAN} components to the residual gives us a linear {\tt REST} (restoration) matrix which both filters the data and interpolates the subtracted foregrounds. 
\begin{equation}
    \RmatRest =  \Amat \left[ \Amat^T \wmat \Amat \right]^{-1} \Amat^T \wmat \left(\Imat - \RmatLazy\right) + \RmatLazy.
\end{equation}\label{eq:DAYENUREST}

We can understand the first term of equation~\ref{eq:DAYENUREST} as follows. First $(\Imat-\RmatLazy)$ is applied which effectively filters out all small-scale structure dominated by the 21\,cm signal and contains RFI flagging gaps. Next, $\Amat^T \wmat$ transforms the flagged data into the DPSS basis. Mode-mixing between the DPSS coefficients, due to flagged channels, is undone by applying $\left[ \Amat^T \wmat \Amat \right]^{-1}$ and a final application of $\Amat$ transforms back into frequency space. Thus, the total action of the first term is the interpolation over flagged channels with fitted smooth DPSS modes. The second term of equation~\ref{eq:DAYENUREST} isolates the fine-frequency components of the signal including noise and the 21\,cm signal itself. 

In \S~\ref{sec:SIMULATIONS}, we will demonstrate the performance of {\tt DAYENUREST} on realistic foreground and signal simulations.

\section{Validation with Realistic Simulations}\label{sec:SIMULATIONS}
In the last section, we tried to understand how demixing and filtering were limited by non-idealities of the signal covariance matrix. To this end, we simulated Gaussian realizations of a simplified foreground model with no consideration of antenna chromaticity or reference to an actual sky with spectral slope. In addition, the dynamic range that we assumed between foregrounds and 21\,cm (eight orders of magnitude in the power-spectrum),  was somewhat 
 less than what is expected for many models. In this section, we validate {\tt DAYENU} by applying it to more realistic simulated visibilities. 

\subsection{Simulation Description}\label{ssec:Simulations}

In this section, we use simulated HERA visibilities \citep[Appendix A,][]{Kern:2019} to validate filtering with $\RmatLazy$ along with the overall impact of this filtering on power-spectrum statistics. 
We construct our simulations using the \texttt{healvis} software \citep{Lanman:2019b}, which integrates the visibility equation using a HEALpix representation of the sky \citep{Gorski:2005}.
The simulations use the Global Sky Model \citep[GSM;][]{deOliveiraCosta:2008} for the foreground model, and a flat-spectrum, uncorrelated random Gaussian field as the EoR model with a variance of 25 mK$^2$.

They also use a simplified model of the HERA primary beam in instrumental XX and YY polarization, assuming minimal frequency structure in the sidelobes of the beam. Specifically, the beam is low-pass filtered across frequency at every HEALpix pixel to reject structures for $|\tau| > 250$ ns.
For this work this is likely an inconsequential feature of the simulations, as it sets at which delay the foreground power dips below the EoR signal, which is not something that our analysis is sensitive to \citep{Fagnoni:2019}.
The simulations span eight hours of local sidereal time (LST) and have a frequency coverage from 120 -- 180 MHz in 256 channels leading to a 235 kHz channelization.
We refer the reader to \citep{Lanman:2019a} for more details on the \texttt{healvis} package and \citep{Kern:2019} for further information on the simulated data products.
Radio frequency interference 
plays a major role in setting the efficacy of these techniques.
In this section, we use flagging masks representative of the RFI environment for HERA's first observing season \citep{Kerrigan:2018, Kern:2020a}.

\subsection{Validating {\tt DAYENU} and {\tt DAYENUREST} as Visibility Filters.}

Aside from being used as a filtering matrix in the final calculation of $\hat{p}_\alpha$, {\tt DAYENU} can readily be employed in sandbox-type data analyses assessing the level of spectral structures in individual visibilities, data-cubes, and other products. In this section, we compare its efficacy to {\tt CLEAN} filtering which is often used to a similar end. To do so, we inspect the performance of the direct application of {\tt DAYENU} and {\tt DAYENUREST} to our simulated visibilities, and compare our results to {\tt CLEAN}. 
In the literature (e.g. \citep{Kern:2019}), {\tt CLEAN}ing is performed on the visibility after zero-padding by $\Ndata$ channels on either side (For these simulations $\Ndata=256$) and taper-filtering with a Tukey window with $\alpha = 0.15$. Zero-padding is performed to give {\tt CLEAN} a larger number of Fourier modes to work with; allowing it to fit the same aharmonic delays that are absent from an $\Ndata$ DFT. We perform {\tt CLEAN}ing over $\pm 150$\,ns in delay-space. Each iteration of {\tt CLEAN} finds the peak power of the data in delay-space and subtracts the peak power times $0.1$ (gain) times a flagging kernel centered at the peak delay until the RMS residual changes with each iteration by less than some fraction of the RMS of the original visibilities. The tolerance parameter can be set as low as we want to obtain some arbitrary degree of foreground subtraction. In practice, the choice of tolerance depends on the constraints of computational resources. We adopt $10^{-9}$ that is currently being used in the HERA analysis pipeline. In addition, for $\Ndata=256$, {\tt CLEAN}ing a single baseline on a single time to $10^{-9}$ tolerance has a similar runtime (within an order of magnitude) of computing the psuedoinverse of $\xcovMatLazy$ to obtain $\RmatLazy$.

\begin{figure*}
    \centering
    \includegraphics[width=\textwidth]{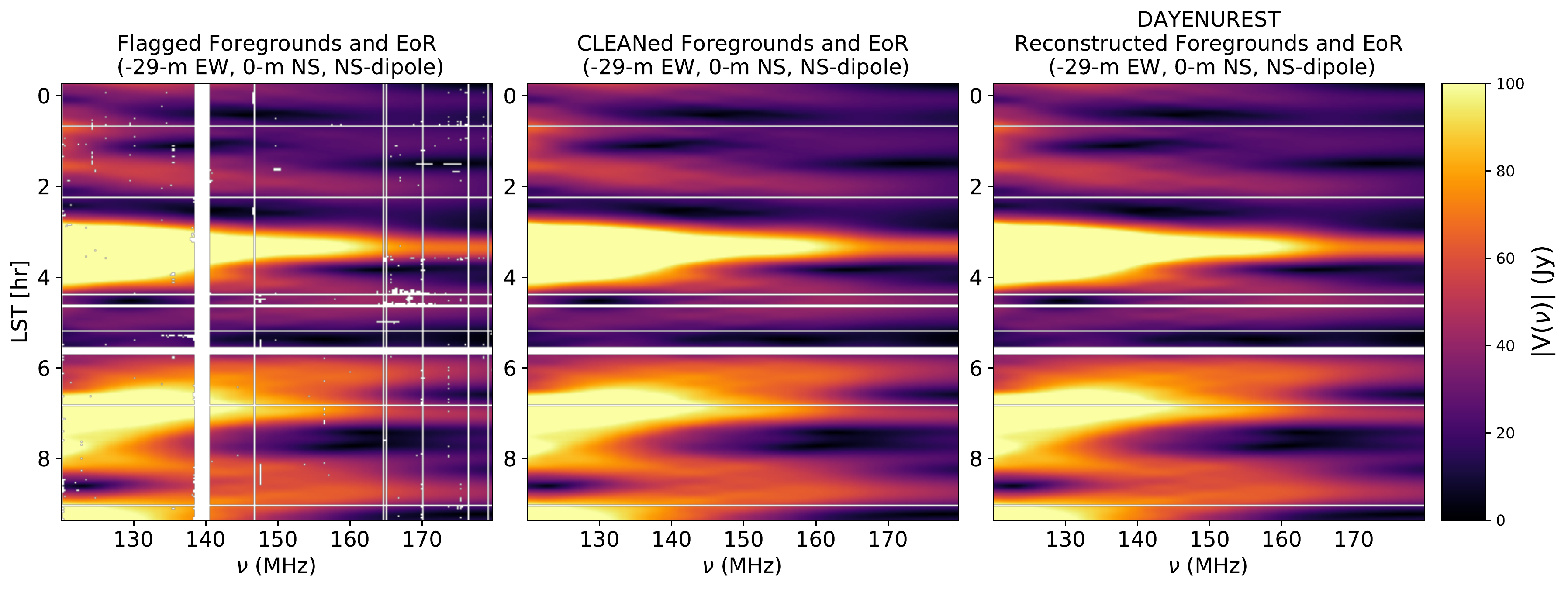}
    \caption{{\bf Left:} A simulated visibility including modeled foregrounds and 21\,cm fluctuations with gaps at the locations of frequency dependent RFI flags. {\bf Center:} Simulated foregrounds and EoR after low-delay frequency interpolation with the {\tt CLEAN} algorithm. {\bf Right:} Simulated foregrounds and EoR after low-delay frequency interpolation with {\tt DAYENUREST}. At the macro-scale, linear in-painting delivers qualitatively similar results to iterative CLEANing. The low-level inconsistences between foreground interpolation by {\tt CLEAN} and {\tt DAYENUREST} are best understood by inspecting the residuals left over after subtracting these foreground models (Figs.~\ref{fig:SIMULATEDRESIDUALS} and \ref{fig:DELAYSPECTRA})}
    \label{fig:SIMULATEDSIGNALS}
\end{figure*}

For ${\tt DAYENUREST}$, we limit the set of DPSS vectors to those with eigenvalues of $\Lmat$ greater then $10^{-12}$. As we stated in \S~\ref{ssec:DPSS}, the maximum eigenvalue of $\Lmat$ is close to unity. We compare the sum of clean residuals and clean components, which interpolate over flagged channel gaps (Center Fig.~\ref{fig:SIMULATEDSIGNALS}), to {\tt DAYENUREST}d simulations (Right Fig.~\ref{fig:SIMULATEDSIGNALS}). At large scales, our linear cleaning and interpolation technique performs just as well as {\tt CLEAN} in reproducing macroscopic foreground features. In order to understand the low-level disagreements between the two, we inspect their residuals.

\begin{figure*}
    \centering
    \includegraphics[width=\textwidth]{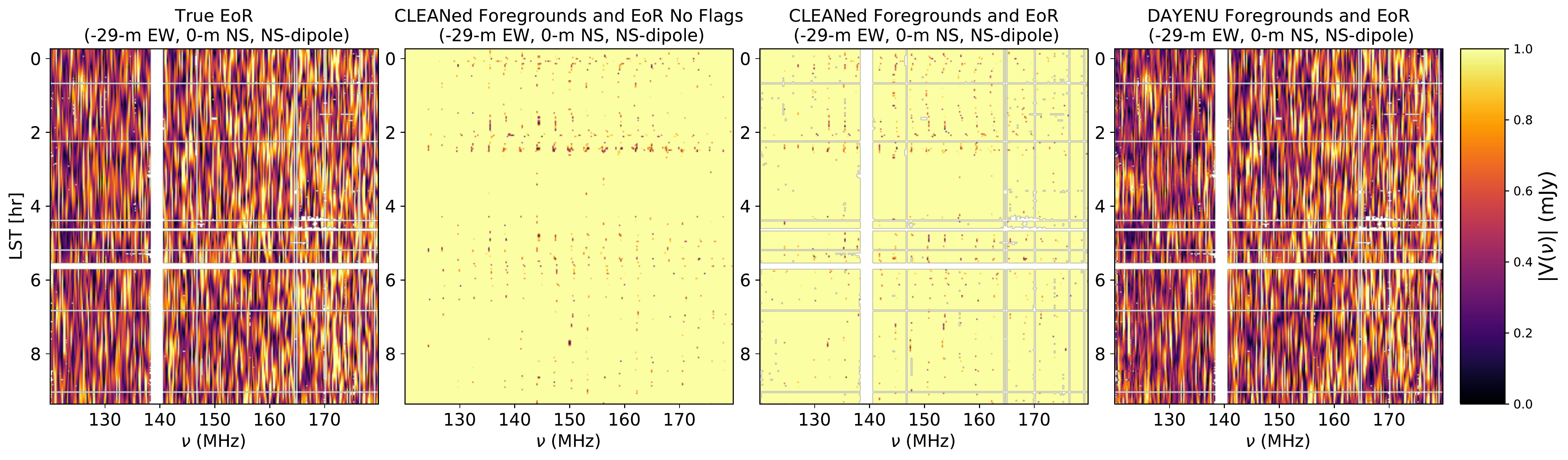}
    \caption{{\bf Left:} an injected mock EoR signal. {\bf Center Left:} Residuals after filtering using the {\tt CLEAN} algorithm with no flagging. {\bf Center Right:} Residuals after foreground filtering using the {\tt CLEAN} algorithm with flagging. The level of real-space {\tt CLEAN} residuals is roughly independent of flagging. Although the {\tt CLEAN} residuals exceed the 21\,cm signal, as long as these residuals are spectrally smooth, they are not an obstacle to detecting 21\,cm in Fourier space. The presence of flagging and residuals presents complications (as we see below in Fig.~\ref{fig:DELAYSPECTRA}).  {\bf Right:} Residuals after foreground filtering using our linear filter. EoR fluctuations remain primarily intact while foregrounds have been completely eliminated.}
    \label{fig:SIMULATEDRESIDUALS}
\end{figure*}

We compare the residuals from {\tt CLEAN} and {\tt DAYENUREST} (Fig.~\ref{fig:SIMULATEDRESIDUALS}). For {\tt CLEAN}, we refer to residuals as what is left in the data after iteratively subtracting all {\tt CLEAN}-components and for {\tt DAYENU} and {\tt DAYENUREST}, as in the previous sections, residuals refer to the data after applying $\RmatLazy$. Note that the residuals for {\tt DAYENU} and {\tt DAYENUREST} are identical by the definition of {\tt DAYENUREST} (eq.~\ref{eq:DAYENUREST}). In Fig.~\ref{fig:SIMULATEDRESIDUALS}, {\tt DAYENU} and {\tt DAYENUREST} subtract the foregrounds to below the 21\,cm level (right panel) while {\tt CLEAN} leaves significant residuals (center right panel). To understand the impact of flagging, we also inspect the residuals of {\tt CLEAN} with no flagging (center left panel). The {\tt CLEAN} residuals are nearly identical whether or not flagging is present. It follows that flagging alone does not impact the absolute level of residuals left after {\tt CLEAN}ing. If these residuals instrinsically stay within the wedge, they will not have an impact on our ability to detect 21\,cm outside of the wedge. However, the presence of flagged channels will cause the residuals to enter the EoR window at a level that depends on the flagging.

\begin{figure*}
    \centering
    \includegraphics[width=\textwidth]{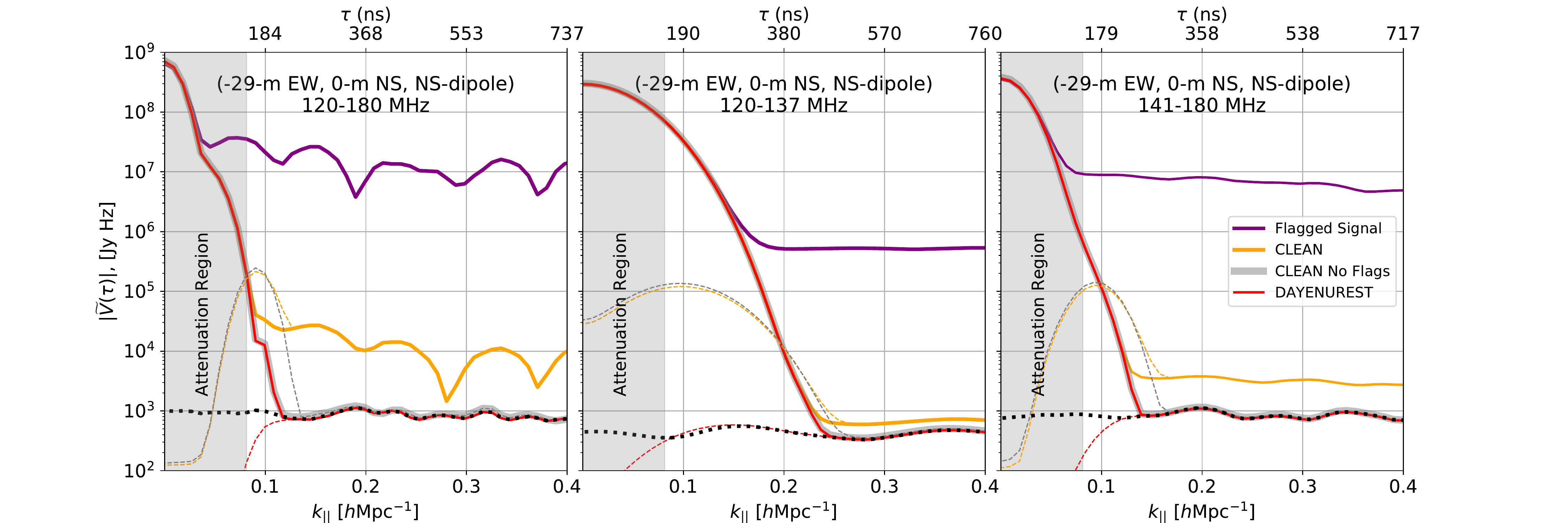}
    \caption{Time averages over eight hours of LST of the absolute value of delay-transformed visibilities in Fig.~\ref{fig:SIMULATEDSIGNALS}, tapered by a 7-term Blackman-Harris window. {\bf Left:} 120-180 MHz (all 256 channels), {\bf Center:} 120-137 MHz (below ORBCOMM), {\bf Right:} 141-179 MHz (above ORBCOMM). Solid lines represent filtered and restored foregrounds and thin dashed lines show residuals. We show the attenuation of our {\tt CLEAN} and {\tt DAYENU} filters as a grey-shaded region.  Over all bands, ringing from RFI flags causes the unfiltered foregrounds (purple lines) to completely mask the 21\,cm signal (black-dotted lines). Pealing and in-painting foregrounds using the {\tt CLEAN} algorithm with a tolerance of $10^{-9}$ leaves significant residuals that exceed the 21\,cm signal in all studied bands and are especially problematic when the FT window includes the heavily flagged ORBCOMM frequencies ($\approx 137$\,MHz). {\tt DAYENUREST} (dashed line) subtracts foregrounds far below the 21\,cm level, allowing for an unbiased estimate of 21\,cm emission outside of the central foreground lobe.}
    \label{fig:DELAYSPECTRA}
\end{figure*}

In Fig.~\ref{fig:DELAYSPECTRA}, we compare the Blackman-Harris taper-filtered delay-transform of {\tt DAYENUREST} and {\tt CLEAN} filtered data with and without flagging across three different bands. For {\tt DAYENUREST} filtered data refers to the data after the application of $\RmatRest$. For {\tt CLEAN} filtered refers to {\tt CLEAN} residuals plus the interpolating {\tt CLEAN} components. Our three bands are as follows. First, the entire 120-180\,MHz band. Second, a 120-138\,MHz band below  ORBCOMM which is heavily flagged, and thirdly $141-180$\,MHz above ORBCOMM with roughly twice the bandwidth as below. With no RFI flagging, {\tt CLEAN} and {\tt DAYENUREST} perform similarly well as can be seen by comparing the red-solid and grey-solid lines in Fig.~\ref{fig:DELAYSPECTRA}. Unfortunately, the presence of RFI flags causes significant bleed of the {\tt CLEAN} filtered data outside of the wedge and is especially bad when the DFT band includes ORBCOMM at $137$\,MHz. We also plot the residuals of {\tt CLEAN} and {\tt DAYENUREST} as dashed lines. The maximum low-delay level of {\tt CLEAN} residuals is practically the same with and without flags. The presence of flags causes these residuals to bleed to high delays at levels much larger then 21\,cm. Since the level of these bleeding residuals agrees with the level of the total filtered data, we conclude that the structures in {\tt CLEAN} residuals introduced by flagging are to blame for high-delay contamination in the {\tt CLEAN} filtered visibilities. Even without ORBCOMM, leakage of {\tt CLEAN} residuals exceeds our injected 21\,cm signal by a factor of a few. {\tt DAYENUREST} (red-solid line) successfully removes foregrounds below the level of the 21\,cm signal (black dotted line) in all cases. The relatively narrow bandwidth below ORBCOMM, presents a potential challenge since the central foregound lobe extends to $k_{\parallel} \approx 0.2 h$Mpc$^{-1}$. Losing $k_\parallel \lesssim 0.2 h$\,Mpc$^{-1}$ to foregrounds has a significant impact on science returns \citep{Pober:2014, EwallWice:2016, EwallWice:2016b}. In \S~\ref{ssec:PSPEC}, we investigate whether the central foreground lobe is actually a fundamental limitation.

Over 256 channels, {\tt CLEAN}'s runtime per integration is also significantly larger than {\tt DAYENUREST}'s. With our adopted parameters, on a laptop with a 2.4\,GHz i5 processor, computing $\RmatLazy$ for each unique flagging pattern and set of filter-widths, centers, and suppression factors takes roughly 0.24 seconds while filtering a baseline at a single time with a cached filter matrix takes approximately $0.003$ seconds. In comparison, the time for {\tt CLEAN} to run on each baseline-time is $0.8$ seconds and there is no possibility of speeding things up through caching.

kBefore we move on to power-spectra, it is worth noting that although we have focused filtering visibilities, $\RmatLazy$ can just as easily be used to foreground-filter gridded visibilites by applying $\RmatLazy$ along the frequency axis of each $uv$ cell. In this situation, one would set $\tau_w$ to include not only the intrinsic chromaticity of the antenna and the wedge in the $uv$ cell but also to include any additional spectral structure that might be introduced by gridding. We leave the question of how much one would need to increase $\tau_w$ for different gridding strategies to future work.

\subsection{Power Spectra}\label{ssec:PSPEC}
\begin{figure}
    \centering
    \includegraphics[width=.5 \textwidth]{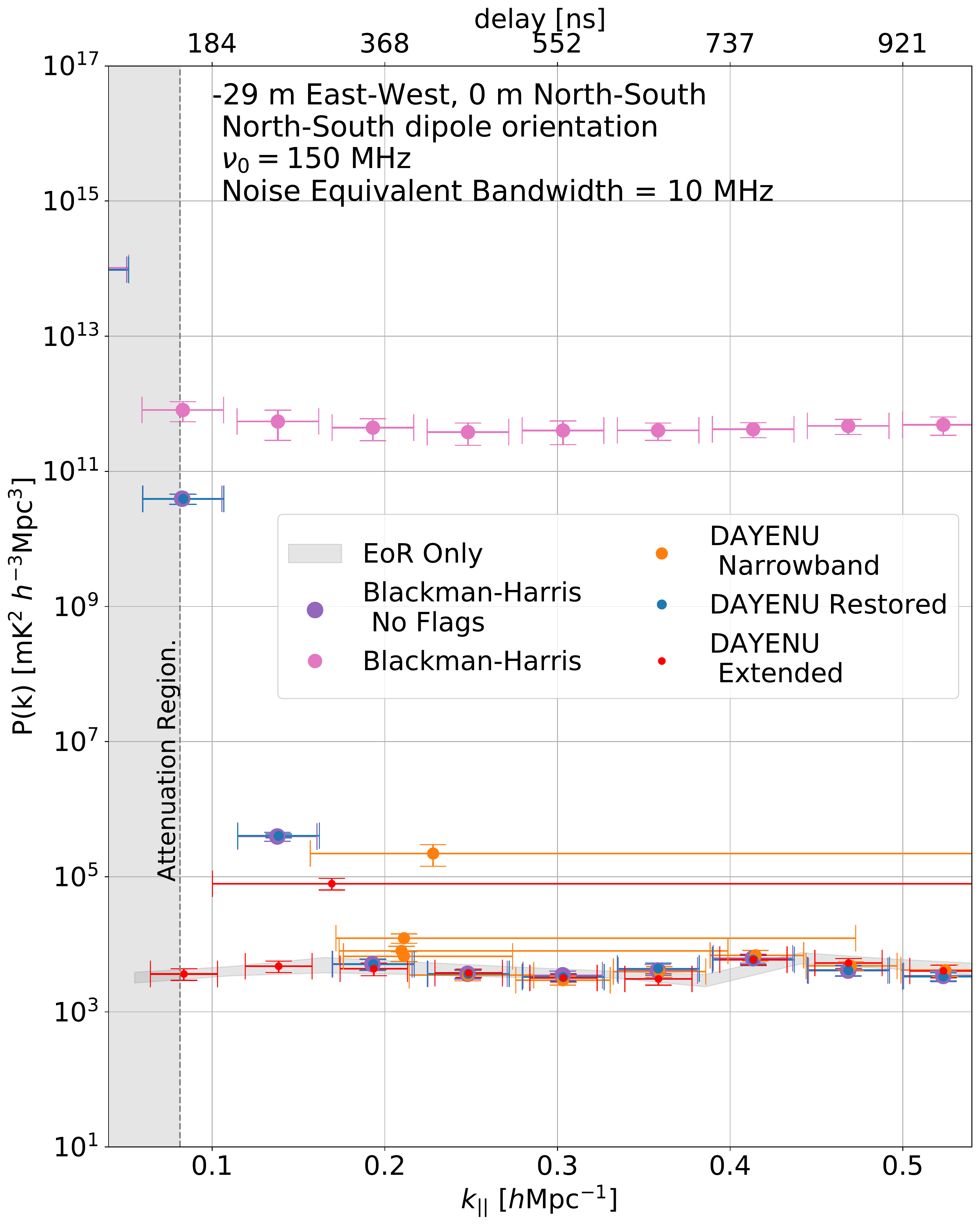}
    \caption{Power spectra estimated from a -29\,m east-west oriented baseline over 10\,MHz noise equivalent bandwidth centered at 150\,MHz and eight hours of LST. Vertical error bars are 68\,\% confidence regions computed from the diagonal of $\SigmaMatHat$ and arise from the sample-variance in 8-hours of sky observations (\S~\ref{sssec:ERRORBARS}). Horizontal errorbars are the 68~\% confidence intervals derived from estimates of the window-function matrix~$\WmatHat$ (\S~\ref{sssec:WINDOWMATRICES}) and points are plotted at 50~\% point of each $\WmatHat$ row. With only a Blackman-Harris apodization filter applied, power-spectrum estimates are heavily contaminated by flagging side-lobes of the foregrounds (pink points). Filtering with {\tt DAYENUREST} and a Blackman-Harris both interpolates the flagged channels and removes power associated with the sharp edges of our finite sample bandwidth (blue points), resulting in a measurement that is in general agreement with an unflagged Blackman-Harris tapered DFT (purple points). Tapered DFT methods that leave the foregrounds in must contend with those foreground's sidelobes. Over 10\,MHz NEB, these sidelobes extend to $\sim 0.2$\,$h$Mpc$^{-1}$, rendering measurements of larger scale modes highly contaminated by foreground bias.  {\tt DAYENU} is a filter that targets and removes foregrounds. But unintentional attenuation of the signal also occurs beyond the edge of the attenuation region (vertical grey filled region) specified by $\tau_w$. If we apply {\tt DAYENU} over 10\,MHz then this attenuation is significant in our single baseline power spectrum out to $0.2$\,$h$Mpc$^{-1}$ (orange points). Applying {\tt DAYENU} across 60\,MHz before estimating our bandpowers from the central 10\,MHz subband allows us to measure bandpowers down to $\sim 0.1$\,$h$Mpc$^{-1}$ with relatively small bias which can be further mitigated using more sophisticated normalization.  }
    \label{fig:POWERSPECTRA}
\end{figure}

We now explore the impact that various choices of $\boldsymbol{\mathsf{R}}$ have on the final power spectrum when when we use identity normalization~$\Mmat \propto \MmatId$. We calculate a normalized ${\bf \hat{p}}$ from 42 channels between 145\,MHz and 155\,MHz; corresponding to a redshift interval of $\Delta z \approx 0.5$ for the following choices of $\Rmat$. 
\begin{itemize}
    \item {\bf Blackman-Harris:} We use an apodization filter with the diagonal set equal to a 7-term Blackman-Harris taper function~$\Rmat = \Rmat^\text{BH}$. To obtain a noise-equivalent bandwidth of 10\,MHz, we extend the spectral window to 96 channels (22.5\,MHz).
    \item {\bf No Flags:} A scenario for reference. The same as Simple Delay-Spectrum but with no RFI flagging. In this scenario, we also have $\Rmat = \Rmat^\text{\bf BH}$
    \item {\bf {\tt DAYENU} Narrowband:} Apply $\RmatLazy$ with $\epsilon=10^{-9}$ and $\tau_w = 150$\,ns across the same bandwidth as the Fourier Transform (42\,channels -- 10\,MHz; $\RmatLazy$). We do not use a taper in the Fourier transform. Thus $\Rmat = \RmatLazy$. 
    \item {\bf {\tt DAYENU} Restored:} Perform linear inpainting of foregrounds using {\tt DAYENUREST} with a 150\,ns attenuation region and in-painting modes spaced by 44.44\,ns ($\RmatRest$). An identical Blackman-Harris tapered Fourier transform as our Blackman-Harris scenario is used to estimate bandpowers from the filtered data. Thus $\Rmat = \Rmat^\text{\bf BH}\Rmat^\text{\bf REST}$. 
    \item {\bf {\tt DAYENU} Extended Filter:} We perform filtering across the entire 60\,MHz band with $\RmatLazy$ before truncating and performing a DFT across the central 10\,MHz. $\Rmat = \RmatLazy$. 
\end{itemize}
In all cases, we use $\Qmatalpha = \Qmatalpha^\text{\bf DFT}$. In order to convert our power spectra from visibility to cosmological units, we multiply $\MmatId$ by a constant 
\begin{equation}
    \Mmat = S \times \MmatId
\end{equation}
where
\begin{equation}
   S = \left(\frac{\lambda^2}{2 k_B} \right)^2 \frac{ X^2Y }{\Ndata^2 \Omega_{pp} B},
\end{equation}
, $\Omega_{pp}$ is the solid angle integral of the primary beam squared and averaged over our band of interest, $Y = d r_\parallel/d \nu$, $X = dr_\perp / d\theta$, $\lambda$ is the average observation wavelength, and $k_B$ is the Boltzmann constant. We refer the reader to \citet{Morales:2004, Parsons:2012b, Parsons:2014} for more the full expressions of these constants and their derivations. We estimate power spectra from eight hours of LST by computing an independent $\phatvec$ every 30.6 seconds and incoherently averaging. Our bandpower estimates appear in Fig.~\ref{fig:POWERSPECTRA} along estimates of vertical and horizontal 68\% confidence errorbars. We derive these confidence intervals from estimates of the bandpower covariances~$\SigmaMatHat$ and window-functions~$\WmatHat$. Before we discuss the results in this plot we first describe our calculations $\SigmaMatHat$ (\S~\ref{sssec:ERRORBARS}) and $\WmatHat$ (\S~\ref{sssec:WINDOWMATRICES}). 

\subsubsection{Error Bars}\label{sssec:ERRORBARS}
\begin{figure*}
    \centering
    \includegraphics[width=\textwidth]{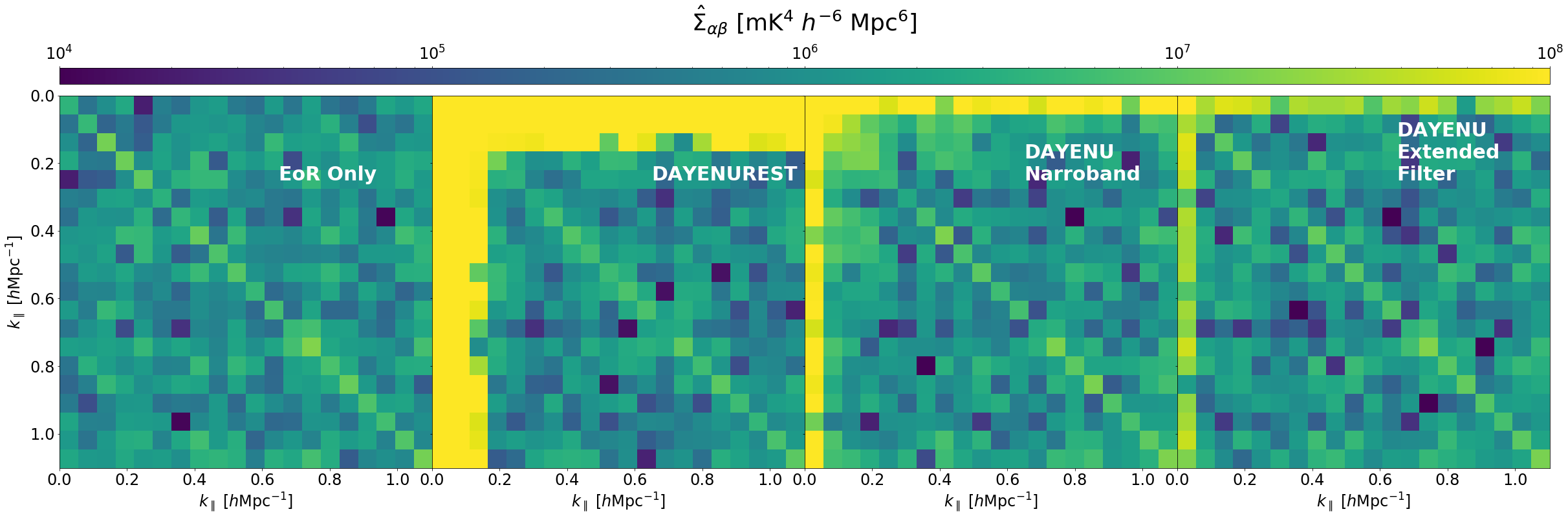}
    \caption{The covariance matrices of $\phatvec$~$\SigmaMatHat$ from which the errorbars in Fig.~\ref{fig:POWERSPECTRA} are derived. {\bf Left:} As a result of flagging, and not attempt to decorrelate, power spectrum errors for the DFT of EoR simulated data are highly correlated. {\bf Center Left:} Errors from {\tt DAYENUREST} which restores foregrounds using linear interpolation with DPSSs and as a result, requires a taper-filtered FT over a larger bandwidth. Error bars are very large below $k_\parallel \lesssim 0.2h$Mpc$^{-1}$ but outside of the foreground region, they are somewhat less correlated then the EoR only panel. This is in part because of the larger DFT and lower side-lobes from a Blackman-Harris. {\bf Center Right:} $\SigmaMatHat$ for {\tt DAYENU} applied over the same 10\,MHz bandwidth of the DFT. Large foreground errors are now contained within the DC bin but significant error correlations exist below $k_\parallel \lesssim 0.2$\,$h$Mpc$^{-1}$.  {\bf Right:} $\SigmaMatHat$ for our {\tt DAYENU} Extended Filtering estimator. Correlations between large $k_\parallel$ modes are similar to the EoR-only and ${\tt DAYENU}$ panels. However, the strong correlations at $k_\parallel \lesssim 0.2$\,$h$Mpc$^{-1}$ that exist when {\tt DAYENU} is applied over a smaller bandwidth have been greatly reduced, as have the foreground errors in the $k_\parallel = 0$\,$h$Mpc$^{-1}$ bin.}
    \label{fig:SIGMAESTIMATE}
\end{figure*}

To calculate $\hat{\sigma}^{\hat{p}}_\alpha$, the standard deviation of our $\alpha^{th}$ bandpower after incoherent averaging, we first calculate $\hat{\sigma}_\alpha^0 \equiv \sqrt{\SigmaMatHatEl_{\alpha\alpha}}$ by empirically computing the covariance of $\hat{p}$ across all LSTs. We show our estimates of $\SigmaMatHatEl$ in Fig.~\ref{fig:SIGMAESTIMATE}. To account for the reduction in errors that occurs from incoherently averaging over the independent realizations of foregrounds and 21\,cm fluctuations in the sky, we use the equation
\begin{equation}
    \hat{\sigma}^{\hat{p}}_\alpha = \hat{\sigma}^0_\alpha \sqrt{\frac{\text{FWHM}^\alpha_c}{T}}
\end{equation}
where $\text{FWHM}^\alpha_c$ is the full-width half-max in time of the correlation between the $\alpha^{th}$ bandpower and itself~$\hat{\Sigma}_{\alpha\alpha}(\Delta t)$  and $T$ is the total amount of time over which LSTs are averaged (8.5 hours). We compute bandpower time-correlations using
\begin{equation}
\hat{\Sigma}_{\alpha\alpha}(\Delta t) = \frac{1}{N_t}\sum_{t} \hat{p}_\alpha(t+\Delta t)  \hat{p}^*_\alpha(t),    
\end{equation}
where $N_t$ is the number of times and $\hat{p}_\alpha(t)$ is the bandpower estimate at each time step. In our case, $N_t=1000$. 
We find the full-width half-max of $\hat{\Sigma}_{\alpha\alpha}(\Delta t)$ using the method {\tt scipy.signal.find\_peaks}. In Fig.~\ref{fig:POWERSPECTRA}, we show the averaged bandpowers and $2$\,$\sigma$ error bars. Since our simulation does not include noise, the errors are purely sourced by sample variance in the foregrounds and signal. 

\subsubsection{Window Matrices}\label{sssec:WINDOWMATRICES}
We estimate window matrices using the equation
\begin{equation}
    \WmatHat = \Mmat \HmatHat,
\end{equation}
where 
\begin{equation}
    \HmatHatEl_{\alpha\beta} = \frac{1}{2} \text{tr} \left( \Rmat^\dagger \Qmatalpha \Rmat \dxcovMatBetaHat \right)
\end{equation}
In practice we do not necessarily have $\HmatHat = \Hmat$ since we don't know the a-priori actual bandpowers of the signal in question and are instead forced to guess some $\dxcovMatBetaHat$. While we technically do potentially have the ability to calculate true bandpowers for our simulated visibilities, we defer an exploration of the consequences of not using true bandpowers to compute $\Hmat$ for paper II. In this paper, we adopt the standard DFT bandpower assumption so that $\dxcovMatBetaHat = \dxcovMatBetaHat^\text{\bf DFT}$. 

\begin{figure*}
    \centering
    \includegraphics[width=\textwidth]{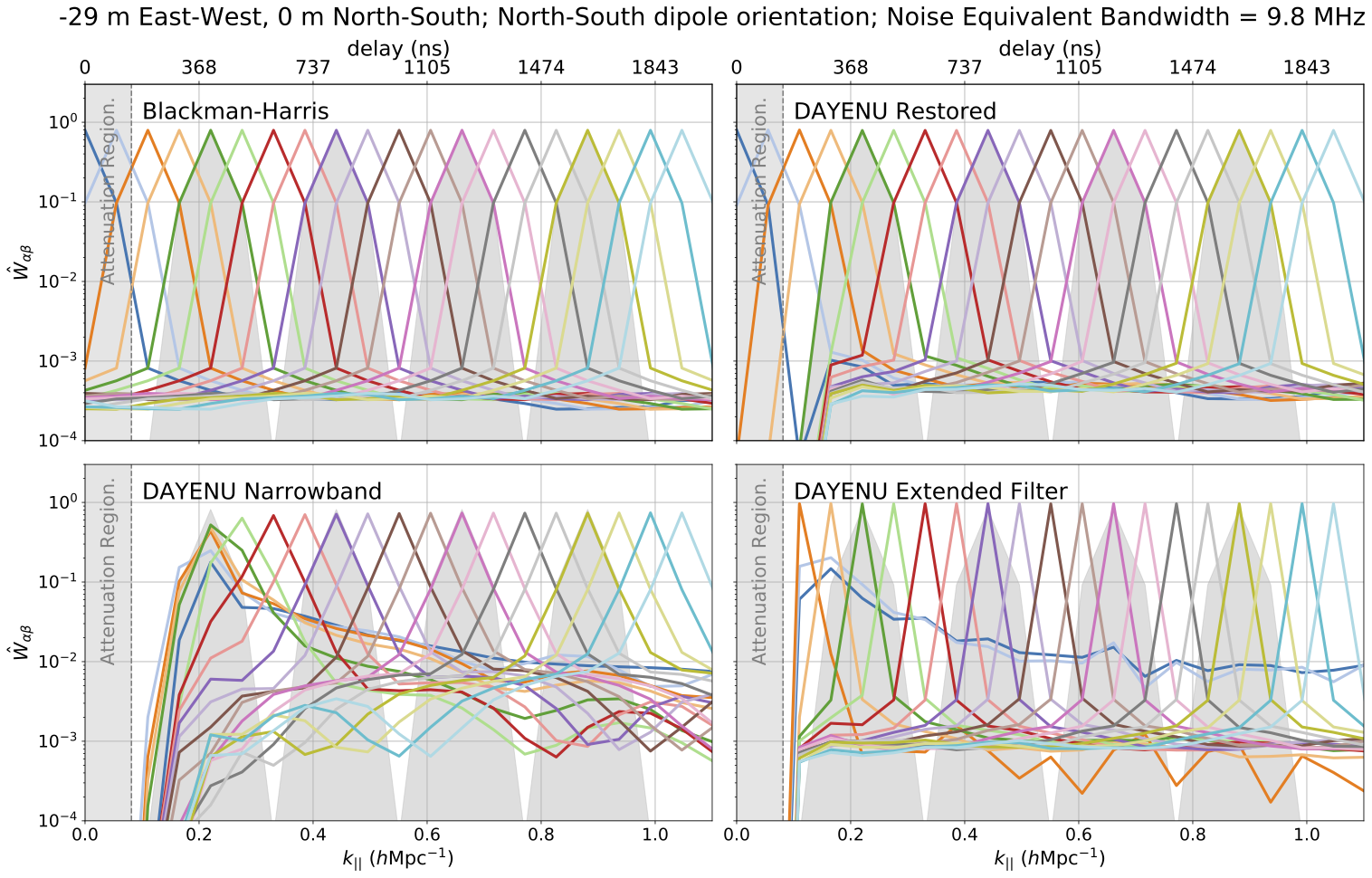}
    \caption{Rows of $\WmatHat$ for the various choices of $\Rmat$ considered in this paper. Each colored line is a different row. We also show every fourth row of $\WmatHat$ for an unflagged Blackman-Harris filtered power spectrum as grey shaded regions. The attenuation set by $\tau_w$ in {\tt DAYENU} is also indicated by a grey shaded region bordered by a dashed line. {\bf Top Left:} Rows of $\WmatHat$ when only a Blackman-Harris filter is used on the flagged visibilities. Window functions exhibit a floor at $\approx -35$\,dB arising from the flags. {\bf Top Right:} When we use the {\tt DAYENUREST} filter, flagging gaps are interpolated over by DPSS vectors that span the attenuation region. This results in the removal of the flagging side-lobes of bandpowers centered within the attenuation region, preventing foreground leakage. Flagging sidelobes remain outside of the attenuation region. {\bf Bottom Left:} Applying {\tt DAYENU} across a narrow band (10\,MHz) removes power within the $150$\,ns attenuation region along with associated side-lobes, eliminating the problem of foreground-flagging sidelobes contaminating all bandpowers. $\WmatHat$ rows that would otherwise be centered inside of the attenuation region are now centered outside and have larger side-lobes that extend to larger wave-numbers. This is because these bandpowers had most of their power eliminated by {\tt DAYENU} but the flagged DFT leaks power back in from high delays. The relatively large amount of unintentional attenuation that accompanies a narrow band filter (see also Fig.~\ref{fig:INJECTIONBANDWIDTH}) prevents us from effectively measuring bandpowers below $k_\parallel \lesssim 0.2h$Mpc$^{-1}$. {\bf Bottom Right:} Our {\tt DAYENU} Extended filter filters over all $60$\,MHz of before performing a DFT over the same  $10$\,MHz as the {\tt DAYENU} and {\tt DAYENU} Restored scenarios. The reduction in unintentional attenuation results in our ability to measure $21\,cm$ fluctuations down to $\sim 0.1$\,$h$Mpc$^{-1}$, enhancing our ability to perform sensitive 21\,cm measurements.}
    \label{fig:WINDOWFUNCTIONS}
\end{figure*}

We show $\WmatHat$ for our various $\Rmat$ choices, averaged over all time-samples, in Fig.~\ref{fig:WINDOWFUNCTIONS}. Our window functions for the Delay Spectrum and DAYENU Restored are very close to each-other outside of the filtering region where they are narrowly peaked but level off at $\approx -35$\,dB. We also plot every fourth row of $\WmatHat$ for an estimator with no flagging and a Blackman-Harris apodization filter in Fig.~\ref{fig:WINDOWFUNCTIONS}. Since these window functions continue to descend below $-35$\,dB, we conclude that the $-35$\,dB floor in most $\WmatHat$ rows is a consequence of flags. In our Blackman-Harris estimator, these -35\,dB side-lobes extend from bandpower estimates inside of the attenuation region just as much as bandpower estimates outside of the attenuation region. If no foregrounds are subtracted, bandpower estimates inside of the attenuation region are heavily contaminated by foregrounds, causing the significant contamination across all bandpowers that we observe in the Blackman-Harris model (pink points) in Fig~\ref{fig:POWERSPECTRA}. Since the vast majority of power within the filtering region is sourced by interpolated and effectively unflagged DPSS modes,the {\tt DAYENU} Restored filter removes the components of side-lobes of bandpowers centered outside of the attenuation region that overlap with the attenuation region. This effectively breaks the coupling of modes outside the attenuation region with the foregrounds. The {\tt DAYENU} Narrowband filter suppresses the coupling of all bandpower estimates with delays inside of the attenuation region and as a consequence, many of the rows of $\WmatHat$ that would typically be centered inside of the attenuation region are now centered at its edge at $k_\parallel \approx 0.2 h$\,Mpc and preventing us from effectively measuring cosmological modes below this value. By extending the filtering bandwidth from $10$ to $60$\,MHz our {\tt DAYENU} Extended filter reduces the width of the attenuation region to $\approx 0.1$\,$h$Mpc$^{-1}$ and allowing for significant improvements in our ability to detect and interpret 21\,cm fluctuations.

\subsubsection{Power Spectrum Results.}
Having explained the source of our vertical and horizontal $68\%$ confidence regions, we dicuss the results of Fig.~\ref{fig:POWERSPECTRA}. The presence of RFI gaps introduces window-function side-lobes at the $-35$\,dB level (Fig.~\ref{fig:WINDOWFUNCTIONS}). Thus, if our $\Rmat$ filter does not attenuate foregrounds before applying $\Qmatalpha^\text{\bf DFT}$, all bandpowers will be heavily contaminated by foregrounds. This is indeed the case for our Blackman-Harris model (pink points). If no flags are present, these flagging side-lobes do not exist and our estimator eventually recovers 21\,cm. However, the smallest $k_\parallel$ that we can access is limited by the Blackman-Harris side-lobes of foregrounds which extend to $k_\parallel \sim 0.2$\,$h$Mpc$^{-1}$. The same is true for the {\tt DAYENU} Restored scenario (blue points). The primary accomplishment of foreground interpolation is to remove the bleed from flagging gaps but we must still contend with the Blackman-Harris sidelobes. {\tt DAYENU} Narrowband (orange points) eliminates foregrounds but also severely attenuates signal out to $\approx 0.2$\,$h$Mpc$^{-1}$. Thus, we are still restricted to $k_\parallel \gtrsim 0.2$\,$h$Mpc$^{-1}$ and samples that would otherwise be foreground contaminated at smaller $k_\parallel$ are instead primarily contributed to by power just outside the attenuation region, leading to the handful of points with very large horizontal error bars piled up at $k_\parallel \approx 0.2$\,$h$Mpc$^{-1}$. By using a larger bandwidth in the filtering step, {\tt DAYENU} Extended reduces the region of excessive attenuation down to $\lesssim 0.1$\,$h$Mpc$^{-1}$ (red points). Hence, by filtering foreground selectively, we can access significantly larger co-moving scales then if we only use apodization tapers. From Fig.~\ref{fig:INJECTIONBANDWIDTH}, we know that our bandpowers are biased low at the $1-10$\,\% level -- something that is technically not significantly detected in our single-baseline analysis due to sample variance errors. However, this bias can have implications for more sensitive spherically binned power spectra.

\section{Conclusions}\label{sec:conclude}

In this paper, we introduced a new method for subtracting foregrounds with a highly approximated inverse covariance filter that we call {\tt DAYENU}. With no flagging, {\tt DAYENU} effectively filters foregrounds using DPSSs which are a set of sequences that maximize power concentration within the wedge. Unlike apodization filters, which subtract power equally from foregrounds and signal, {\tt DAYENU} targets and subtracts low-delay foregrounds with minimal impact on high delay signal and noise. {\tt DAYENU} avoids the band edge signal attenuation that is a feature of multiplicative taper filters. 
{\tt DAYENU} is fast, only requiring that one take the psuedo-inverse of a modestly-sized analytic covariance for each baseline length and unique flagging pattern while its linearity allows us to propagate its effect into error estimates and other statistical calculations. We have tested {\tt DAYENU} on simulated visibilites, but in principal it can also filter foregrounds from gridded $uv$ data by applying it to each $uv$ cell instead of each baseline provided that $\tau_w$ is increased sufficiently to include gridding artifacts. Applying {\tt DAYENU} to realistic simulations, we have learned the following: 
\begin{enumerate}
    \item {\tt DAYENU} is effective at subtracting delay-limited foregrounds at the $\lesssim 10^{-6}$ level, even in the presence of significant flagging (Figs.~\ref{fig:AUTOFIRSTLOOK} and \ref{fig:SIMULATEDRESIDUALS}). If applied across a $\approx 100$\,MHz band, signal attenuation is kept below $\approx 1\%$ beyond 300\,ns of the delay-space filter edge. This attenuation can be corrected further in the power-spectrum normalization step. {\tt DAYENU}'s efficacy over filtering with a DFT arises from the fact that, unlike the DFT, it down-weights foreground wedge structures that are not harmonices of $B^{-1}$. 
    
    \item A combination of {\tt DAYENU} and least-squares fitting of DPSSs ({\tt DAYENUREST}) is a fast, linear alternative to the iterative {\tt CLEAN} algorithm whose residuals are significantly smaller than {\tt CLEAN}'s given similar computing times (Figs.~\ref{fig:SIMULATEDSIGNALS} and \ref{fig:SIMULATEDRESIDUALS}). 
    
    \item Applying {\tt DAYENU} across a $\sim60-100$\,MHz band before estimating bandpowers over the $\sim10$\,MHz necessary for stationary 21\,cm statistics allows us to access LoS scales of $\lesssim.15\,h$Mpc$^{-1}$ that, even without flagging, are inaccessible to apodized DFTs  (Fig.~\ref{fig:POWERSPECTRA}) and (Fig.~\ref{fig:WINDOWFUNCTIONS}). 
    
\end{enumerate}
Our takeaway from examining {\tt DAYENU} is that in the regime where baselines are short so that their information is mutually independent, an inverse covariance filter that is good enough for us is simply one that captures the large dynamic range between foregrounds and signals over the wedge delays and includes information on the frequency structures in the the foreground wedge that are not harmonics of $B^{-1}$. We have shown that a simple covariance like $\RmatLazy$ can be many orders of magnitude different from that of the true data covariance but still serve as a highly effective filter. This bodes well for 21\,cm and other intensity mapping applications where the precision characterization of our instruments and foregrounds is difficult.

\section*{Code}

An interactive jupyter tutorial on using DAYENU can be found at \url{https://github.com/HERA-Team/uvtools/blob/master/examples/linear_clean_demo.ipynb}. DAYENU's source code can be found at \url{https://github.com/HERA-Team/uvtools/blob/master/uvtools/dspec.py}

This work made use of the {\tt numpy} \citep{Scipy:2020} , {\tt scipy} \citep{Scipy:2020}, {\tt matplotlib} \citep{Matplotlib:2007}, {\tt aipy} \url{https://github.com/HERA-Team/aipy}, and {\tt astropy} \url{https://www.astropy.org/} and {\tt jupyter} \url{https://github.com/jupyter/jupyter} python libraries along with {\tt pyuvdata} \citep{Hazelton:2017} and {\tt healvis} \citep{Lanman:2019b} python packages. 

\bibliographystyle{mnras}
\bibliography{Dayenu}

\section*{Acknowledgements}

We thank Jacqueline Hewitt, Honggeun Kim, Kevin Bandura, Miguel Morales, Bobby Pascua, Bryna Hazelton, and Ue-Li Pen for helpful discussions. 
AEW and acknowledges support from the NASA Postdoctoral Program and the Berkeley Center of Cosmological Physics. JSD gratefully acknowledges the support of the NSF AAPF award \#1701536. A portion of this work was carried out at the Jet Propulsion Laboratory,
California Institute of Technology, under a contract with the National
Aeronautics and Space Administration. 
AL acknowledges support from the New Frontiers in Research Fund Exploration grant program, a Natural Sciences and Engineering Research Council of Canada (NSERC) Discovery Grant and a Discovery Launch Supplement, the Sloan Research Fellowship, as well as the Canadian Institute for Advanced Research (CIFAR) Azrieli Global Scholars program.
This material is based upon work supported by the National Science Foundation under grants \#1636646 and \#1836019 and institutional support from the HERA collaboration partners. 
This research is funded in part by the Gordon and Betty Moore Foundation. 
HERA is hosted by the South African Radio Astronomy Observatory, which is a facility of the National Research Foundation, an agency of the Department of Science and Innovation.

\appendix

\section{The Dependence of {\tt CLEAN} residual amplitudes on the tolerance parameter. }
\begin{figure}
    \centering
    \includegraphics[width=.5\textwidth]{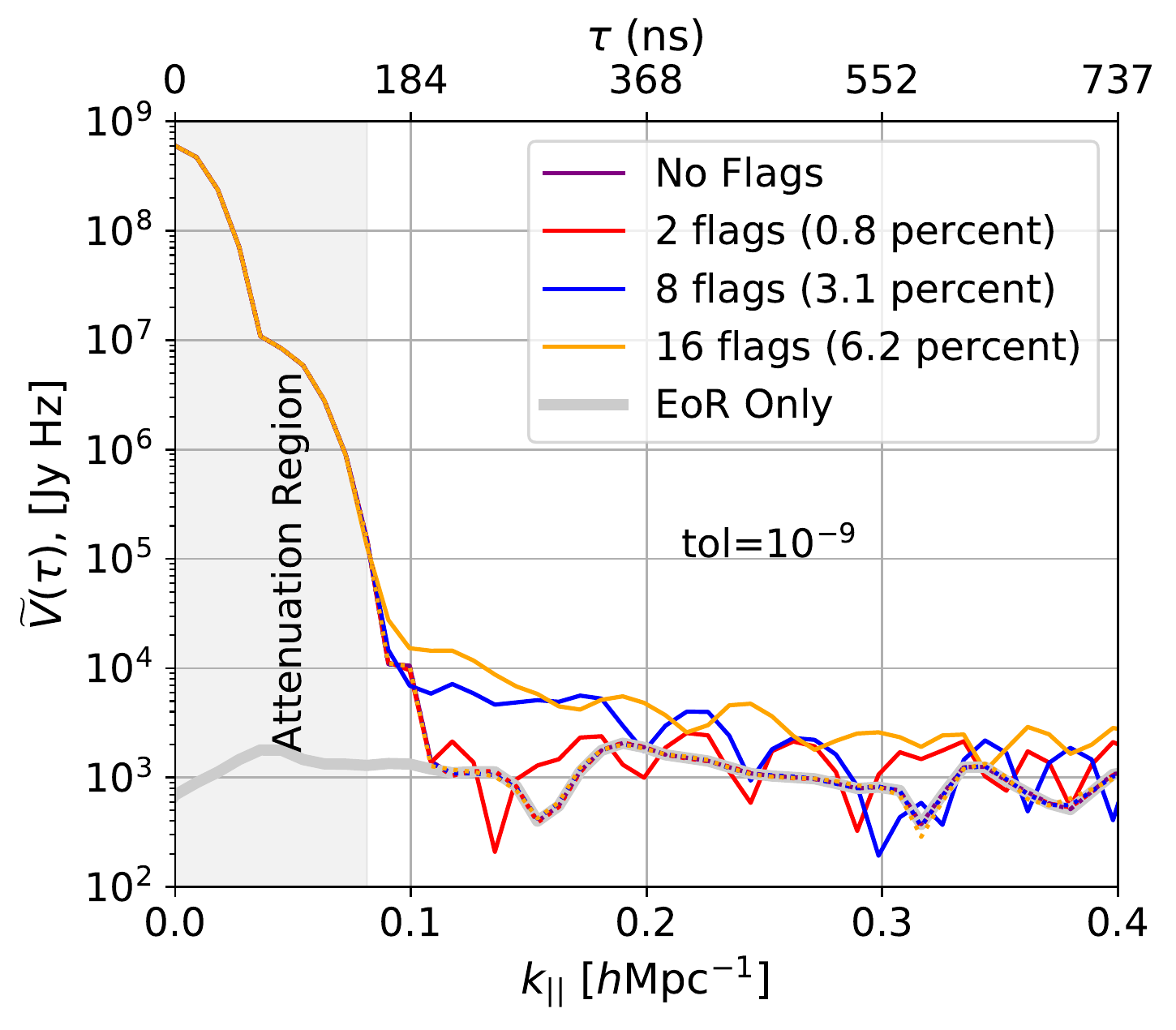}
    \includegraphics[width=.5\textwidth]{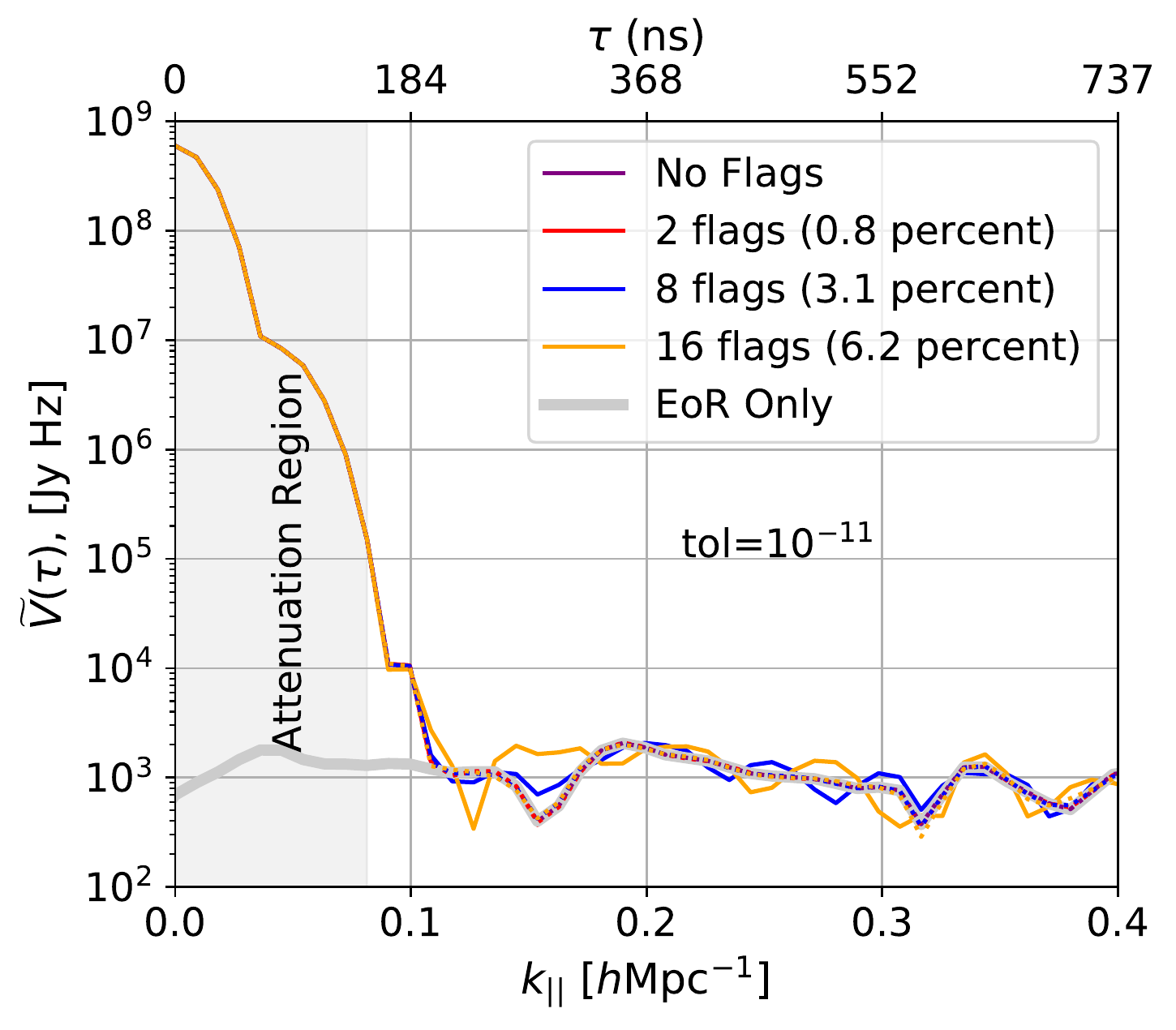}
    \caption{Top: Delay-transformed {\tt CLEAN}ed visibilites for {\tt tol}=$10^{-9}$ (top panel) and {\tt tol}=$10^{-11}$ (bottom panel). Different colors denote different numbers of contiguous flagged channels centered at the $137$\,MHz ORBCOMM frequency. No other flags are introduced and {\tt CLEAN} is performed over the entire band. Dotted lines are the results of applying {\tt DAYENU} to the various levels of flagging. The {\tt DAYENU} filtered visibilities are in very good agreement with the signal outside of the attenuation region. }
    \label{fig:CLEAN_FLAGGING_PERFORMANCe}
\end{figure}

In our comparison, we assumed a fixed set of {\tt CLEAN} parameters employed by the HERA pipeline \citep{Kern:2019} and the RFI environment of the Karoo radio observatory. The presence of flagging leaks residuals left over by {\tt CLEAN}ing across all delays. Hampering a 21\,cm detection. Lowering the residuals also lowers this leakage so in principal decreasing the tolerance should allow for sufficiently low residuals for a 21\,cm detection. In this appendix, we examine the {\tt CLEAN} performance as a function of flagging percentage and tolerance parameter. We run {\tt CLEAN} for a single model baseline and time across all 256 channels with 256 channel zero-padding on either side and a Tukey taper. We iteratively increase the width of flagging on the ORBCOMM band; starting with no flags, then introducing two 235\,kHz channels centered at 137\,MHz. Next, we introduce four channels, eights channels, and sixteen channels. In the top-panel of  Fig.~\ref{fig:CLEAN_FLAGGING_PERFORMANCe}, we compare residuals for different levels of flagging to the injected 21\,cm signal. Even when two channels are flagged, significant deviations are introduced in {\tt CLEAN} when the tolerance is set to $10^{-9}$ (solid colored lines). On the other hand, {\tt DAYENUREST} reproduces both the foregrounds and signal with no residual bias. 

As we mentioned above, the biases from {\tt CLEAN} arise from foreground residuals that have not been fully subtracted and still contain side-lobes from flagging. By decreasing the {\tt tol} parameter in {\tt CLEAN}, we can actually subtract deeper. Thus, in principal there should exist small enough values OF THE tolerance such that side-lobes are suppressed enough to recover 21\,cm fluctuations without signifant foreground bias. We explore this possiblity by lowering the tolerance to $10^{-11}$ (Fig.~\ref{fig:CLEAN_FLAGGING_PERFORMANCe} bottom-panel). Given this lower value, residuals are not visibly present with two flagged channels but $\gtrsim 10\%$ biases appear after $\gtrsim 8$ channels (only 3.1\% of the data) are flagged. Running {\tt CLEAN} with {\tt tol}=$10^{-11}$ takes 22 seconds per baseline and time-sample on a 2.4\,GHz i5 processor -- $\sim 100$ times slower then the linear filter if $\RmatLazy$ is computed at every baseline time and $ \sim 10^4$ times slower then the realistic scenario where all baseline-times can be filtered with cached matrices. 

While decreasing the tolerance can lower foreground leakage, there are diminishing returns and even after a $10^4$ performance hit relative to {\tt DAYENU}, we run into trouble with just $3\%$ of channels flagged. 

\bsp	
\label{lastpage}

\end{document}